\documentclass{lmcs} %%% last changed 2014-08-20
\pdfoutput=1
\usepackage[utf8]{inputenc}

\usepackage{lastpage}
\lmcsdoi{22}{2}{23}
\lmcsheading{}{\pageref{LastPage}}{}{}%
{Jun.~25,~2025}{May~25,~2026}{}


\usepackage{hyperref}
\usepackage[linesnumbered,plainruled]{algorithm2e}
\usepackage{soul}
\usepackage{float}
\usepackage{xcolor}
\usepackage{tikz}
\usepackage[normalem]{ulem}
\usetikzlibrary{fit,calc}
\usetikzlibrary{positioning}
\usetikzlibrary{arrows.meta}

\theoremstyle{plain} %\crefname{satz}{Satz}{S\"atze}
\def\crypto{{\sc CryptoMiniSat}}
\def\xmaple{{\sc xMapleLCM}}
\def\xmapleBold{{\textbf{\textsc{xMapleLCM}}}}
\def\maple{{\sc MapleLCM}}
\def\kissat{{\sc Kissat}}
\def\sbva{{\sc SBVA-CaDiCaL}}
\def\glucoser{{\sc GlucosER}}

\newcommand{\reviewOut}[1]{}

\begin{document}

% If the title is longer than 55 characters, then specify a shorter running title as the optional argument to \title. The running title should be roughyl at most 55 characters:
\title[Extended Resolution Clause Learning via DIPs]{Extended Resolution Clause Learning via Dual Implication Points}
%% \titlecomment{{\alsuper*}OPTIONAL comment concerning the title, \eg,
%%   if a variant or an extended abstract of the paper has appeared elsewhere.}
\thanks{Albert Oliveras is supported by grant PID2021-122830OB-C43, funded by MCIN/AEI/ 10.13039/501100011033 and by “ERDF: A way of making Europe”; by grant PID2024-157044OB-C32, funded by MICIU /AEI /10.13039/501100011033 / FEDER, UE. and by Barcelogic through research grant C-11423.}	%optional

% affiliations are numbered automatically with a, b, c (see below)
% use the optional argument to indicate the affiliation(s) of each author
% omit the argument if there is only one author, or only one affiliation
\author[S.~Buss]{Sam Buss\lmcsorcid{0000-0003-3837-334X}}[a]
\author[J.~Chung]{Jonathan Chung\lmcsorcid{0000-0001-5378-1136}}[b]
\author[V.~Ganesh]{Vijay Ganesh\lmcsorcid{0000-0002-6029-2047}}[c ]
\author[A.~Oliveras]{Albert Oliveras\lmcsorcid{0000-0002-5839-1911}}[d]

% affiliation 1 (automatically numbered a)
\address{University of California, San Diego, La Jolla, CA, USA}
% write emails for all authors having that affiliation
%email{name1@email1, name2@email1, name3@email1}  %optional

% affiliation 2 (automatically numbered b)
\address{Lorica Cybersecurity, Canada}
%\email{name2@email2}  %optional

% affiliation 3 (automatically numbered c)
\address{Georgia Institute of Technology, Atlanta, USA}
%\email{name2@email2}  %optional

% affiliation 4 (automatically numbered d)
\address{Universitat Politècnica de Catalunya, Barcelona}
%\email{name2@email2}  %optional

%% etc.

%% required for running head on odd and even pages, use suitable
%% abbreviations in case of long titles and many authors:

%%%%%%%%%%%%%%%%%%%%%%%%%%%%%%%%%%%%%%%%%%%%%%%%%%%%%%%%%%%%%%%%%%%%%%%%%%%

%% the abstract has to PRECEDE the command \maketitle:
%% be sure not to issue the \maketitle command twice!

\begin{abstract}
  \noindent
  We present a new extended resolution clause learning (ERCL) algorithm, implemented as part of a conflict-driven clause-learning (CDCL) SAT solver, wherein new variables are dynamically introduced as definitions for {\it Dual Implication Points} (DIPs) in the implication graph constructed by the solver at runtime. DIPs are generalizations of unique implication points and can be informally viewed as a pair of dominator nodes, from the decision variable at the highest decision level to the conflict node, in an implication graph. We perform extensive experimental evaluation to establish the efficacy of our ERCL method, implemented as part of the {\maple} SAT solver and dubbed {\xmaple}, against several leading solvers such as {\kissat}, {\crypto}, and {\sbva}, the winner of SAT Competition 2023. We show that {\xmaple} outperforms these solvers on Tseitin, XORified formulas and Interval Matching problems. We further compare {\xmaple} with {\glucoser}, a system that implements extended resolution in a different way, and provide a detailed comparative analysis of their performance.
\end{abstract}

\maketitle

\section{Introduction}
\label{sec:introduction}
Over the last several years, Conflict-Driven Clause-Learning (CDCL) SAT solvers have had a dramatic impact on many fields including software engineering~\cite{HandbookSATSoftwareVerification}, security~\cite{xie2005security}, and AI~\cite{satplan1,satplan2}. As solvers continue to be adopted in increasing complex settings, the demand for greater efficiency and reasoning power by users continues unabated. 

While developers continue to improve CDCL SAT solvers, it is simultaneously true that the base solvers are provably no more powerful than the relatively weak general resolution (Res) proof system~\cite{AFT11ClauseLearning,PipatsrisawatDarwiche:modernclauselearning}, and therefore are fundamentally limited. Hence, solver developers have been actively researching novel algorithms that implement stronger proof systems that go beyond Res. Examples of such algorithms include satisfaction-driven clause-learning (SDCL) SAT solvers~\cite{HKB:RedundancySDCL,SDCLMaxSAT}, bounded variable addition (BVA)~\cite{BVA}, symmetry breaking~\cite{HandbookSATSymmetry}, and extended resolution (ER) solvers such as {\glucoser}~\cite{Audemard2010ARO}.

Continuing this trend of strong proof system implementations, we present a new extended resolution clause learning (ERCL) algorithm, incorporated into a CDCL SAT solver, where new variables are dynamically introduced as definitions for {\it dual implication points} (DIPs) in the implication graph constructed by the solver at run time. The concept of a DIP is best understood as a generalization of unique implication points (UIPs). Informally, a UIP can be defined as a dominator node in an implication graph, corresponding to a variable at the highest decision level (DL), that {\it dominates} all paths from the decision variable node at the highest DL to the conflict node. By contrast, a DIP is a pair of dominator nodes in an implication graph such that any path from the decision variable node at the highest DL to the conflict node must pass through at least one node in the pair\footnote{While it is natural to generalize the concept of a DIP to K-Implication Points or $k$-IPs, we do not discuss them in this paper.}.

Implementation of the ERCL algorithm requires several additional methods. First, we need a method to identify DIPs, i.e., a technique that takes as input an implication graph and outputs a DIP and does so in time linear in the size of the input. Second, we need a technique replaces this DIP pair with a new variable and appropriately modifies the clause learning algorithm to learn new clauses involving DIPs. Third, we need an ER framework, built on top of a CDCL SAT solver, that enables new variable addition, ER clause addition and deletion, etc. Finally, we need heuristics that specialize the above mentioned methods in a variety of ways, such as clause learning and clause deletion policies that are based on different kinds of DIPs. We implement all of these methods as part of the MapleLCM solver~\cite{MapleLCM}, and refer to the resulting solver as {\xmaple}. In fact our proposed ERCL method, and its implementation, is very general and easily extensible thus encouraging future exploration and specialization efforts with a variety of heuristics.

\bigskip
\noindent
{\bf Contributions.}
\medskip

\begin{enumerate}
    \item {\bf DIP:} We introduce the concept of dual implication points (DIP), a generalization of UIPs in conflict graphs. We also came up with an algorithm that computes them in linear time. However, due to space limitations and to the non-trivial nature of the procedure, we discuss this in a separate paper~\cite{BGO:TVBs}.

      \medskip
    \item {\bf ERCL Algorithm:} We introduce a highly parameterizable DIP-based ERCL algorithm. The existence of a multitude of different DIPs in a single conflict graph allows us to derive a large variety of ERCL algorithms. This flexibility is crucial in adapting the procedure to different scenarios, unlike previous methods that couple CDCL with extended resolution.

      \medskip
    \item {{\xmapleBold}:} We present a highly extensible and general ER framework as part of {\xmaple}, which allows developers to easily add their own new variable addition, ER clause learning/deletion, and branching policies. Given that DIP-based ERCL is highly flexible by nature, such a framework is necessary to quickly prototype new procedures.

      \medskip
    \item {\bf Experiments:} We perform extensive empirical evaluation and ablation studies on four different classes of instances, namely, SAT Competition 2023 Main Track, random $k$-xor, Tseitin, and Interval Matching, and compare {\xmaple} against leading solvers such as {\kissat}~\cite{kissat}, {\crypto}~\cite{cryptominisat}, {\sbva}~\cite{SBVA}, and the extended-resolution solver {\glucoser}~\cite{LBDAudemardSimon}. Results show that on the last three sets of hard combinatorial formulas, CDCL SAT solvers perform very poorly, whereas both {\xmaple} and {\glucoser}  present much better performance.
      This demonstrates that extended resolution can be added to CDCL and improve the performance of these solvers; {\glucoser} and our DIP-based methods give two different ways to achieve this.
      Moreover, for our DIP-based learning, a simple heuristic  allows us to detect on-the-fly whether extended resolution is being helpful and revert to using standard CDCL in order to get the best of both worlds. This technique enables {\xmaple} to perform similarly to MapleLCM on the SAT Competition 2023 Main Track instances.        
\end{enumerate}

% \begin{enumerate}
%     \item We introduce the concept of dual implication points (DIP) in conflict graphs. We also came up with a method to compute DIPs in linear time, discussed in a separate paper due to space limitations~\cite{BGO:TVBs}. As we discuss below, DIPs are a generalization of UIPs.

%     \item We introduce a DIP-based ERCL algorithm. DIPs provide a flexible way to choose variables to introduce by ER within the traditional CDCL SAT solving framework.
    
%     \item We present a highly extensible and general ER framework as part of xMapleLCM, which allows developers to easily add their own new variable addition, ER clause learning, ER clause deletion, and branching policies.

%     \item We perform extensive empirical evaluation and ablation studies on 4 different classes of instances, namely, SAT Competition 2023 Main Track, random-k-XOR, Tseitin, and interval matching, and compare xMapleLCM against leading solvers such as Kissat 3.1.1~\cite{kissat}, Cryptominisat 5.11~\cite{cryptominisat}, SBVA Cadical~\cite{SBVA} (winner of the SAT Competition 2023 Main Track), and the extended-resolution solver {\glucoser}~\cite{LBDAudemardSimon}. Our results show that xMapleLCM outperforms all but one solver, namely, {\glucoser}, on hard combinatorial instances such as Tseitin, random-k-XOR, and interval matching. 
% \end{enumerate}

\section{Related Work}
\label{sec:related}
The idea of using ER in SAT solving has been studied in various forms in the literature for nearly two decades. The closest approach to ours is {\glucoser}~\cite{Audemard2010ARO}, where extended variables are introduced dynamically during the CDCL search: whenever two consecutive learned lemmas are of the form $\lnot l_1 \lor C$ and $\lnot l_2 \lor C$, with $l_1$ and $l_2$ being their UIPs, then an extended variable $z \leftrightarrow l_1 \lor l_2$ is generated and any future lemma of the form $l_1 \lor l_2 \lor D$ is replaced by $z \lor D$. Hence, information across different conflict analysis steps is used to detect which extended variables to introduce.

Our method differs significantly from {\glucoser} in the way extended variables are identified: we choose DIPs, which can be seen as pairs of variables for which adding a definition would create a better first UIP (often written as 1UIP) in the conflict graph, whereas {\glucoser} definitions are constructed using already existing UIPs. Our method focuses on a single conflict analysis step to extract an extended resolution variable, whereas {\glucoser} uses two steps. Also, unlike in {\glucoser}, our approach does not always learn the standard 1UIP clause, but multiple clauses might be learned that take into account the newly introduced variable. On the other hand, there are similarities at the heart of both procedures: a certain restriction of ER is considered and introduced variables are used to shorten subsequently found clauses.

Another related procedure is presented in~\cite{Huang:ExtRes}. For a learned clause $C$ of size $>2$, the author suggests to split $C$ into $\alpha \lor \beta$, where  $|\alpha| \geq 2$ and $\beta$ is non-empty.
Instead of learning $C$, the solver learns $x \lor \beta$ and $x\leftrightarrow \alpha$, where $x$ is a fresh variable. Experimental results in~\cite{Huang:ExtRes} are limited, and no implementation is available.

More recent uses of ER in SAT solvers are Bounded Variable Addition (BVA)~\cite{BVA} and its structured version SBVA~\cite{SBVA}, that due to clever strategies are able to identify new extended variables whose introduction can reduce the number of clauses. Essentially, the ultimate goal of BVA techniques is to reduce the size of the formula by reducing the number of clauses at the cost of adding a new variable. One big difference  with our work is that this process is done only in a preprocessing step. Finally, one additional direction that has been researched is the development of a BDD-based solver to generate ER proofs~\cite{ExtendedResolutionBDDs}.

Other approaches aiming at improving CDCL solvers by allowing them to use a more powerful proof system are related to the Propagation Redundancy notion~\cite{ShortProofs,StrongExtensionFree}, either via preprocessing steps~\cite{preprocessingPR} or  via the use of the SDCL algorithm~\cite{PruningThroughSatisfaction,SDCLMaxSAT}. While these methods implement proof systems that are stronger than Res, many of them (without new variable addition) are known to be weaker than ER. 

  Another related work is~\cite{PD08Learning}, where the authors introduce a new conflict analysis procedure that learns lemmas that contain two literals of the last decision level. These two literals are indeed a DIP. 
  However, there can be many other candidate DIP's that are not detected by the methods of~\cite{PD08Learning}.  Furthermore, as we will show in Section~\ref{sec:algorithm}, 
  their procedure might sometimes fail to detect any DIPs at all. An additional, important differences w.r.t.\ our work is that \cite{PD08Learning} do not introduce a new variable representing the conjunction of the DIP members. 
  Hence, this is not an extended resolution-based approach and, from the proof-complexity point of view, is not more powerful than standard CDCL or general resolution.

\section{Preliminaries}
\label{sec:preliminaries}

We assume that the reader is familiar with the satisfiability (SAT) problem and the CDCL algorithm, and we refer her to the Handbook of Satisfiability for an excellent overview of these topics~\cite{HandbookCDCL}. Below we focus on conflict analysis, which is the most relevant ingredient from the CDCL algorithm for this paper. We do so by means of the following example.

\begin{exa}
    \label{ex:conflict}
    Consider the following clauses
    $$\begin{array}{llllll}
        (1) & y_1 \lor \lnot x_1 \lor x_2 & (6) & \lnot x_5 \lor x_7     & \!\!(11) &\!\!\!\!  \lnot x_{11} \lor x_{12}\\
        (2) & \lnot x_1 \lor \lnot x_3                & (7) &  x_6 \lor \lnot x_7 \lor x_8 & \!\!(12)\!\!\!\! & x_{10} \lor \lnot x_{11} \lor x_{13}\\
        (3) & y_2 \lor \lnot x_1 \lor x_4 & (8) & \lnot y_3 \lor \lnot y_4 \lor \lnot x_5  \lor \lnot x_9 &\!\! (13)\!\!\!\! & x_{12} \lor \lnot x_{13}\\
        (4) & \lnot y_3 \lor \lnot x_2 \lor x_3 \lor \lnot x_4 \lor x_5 & (9) & \lnot y_4 \lor x_{9} \lor \lnot x_{10} \\
        (5) & y_1 \lor \lnot x_5 \lor \lnot x_6 & (10) & \lnot y_5 \lor y_6 \lor \lnot x_8 \lor x_9 \lor x_{11}\\
        \end{array}
    $$

    \medskip
Assume that CDCL has constructed an assignment that contains, among others, literals $\{\lnot y_1,\lnot y_2, y_3, y_4, y_5, \lnot y_6\}$. Since no propagation is possible, it now decides to add the decision literal~$x_1$. Due to clause (1), we can unit propagate literal $x_2$, being (1) the reason of $x_2$ and its antecedents $\{\lnot y_1,x_1\}$. Similarly, $\lnot x_3$ is propagated due to reason (2), with antecedents $\{x_1\}$, and $x_4$ due to reason (3) with antecedents $\{\lnot y_2, x_1\}$. If we continue this process we eventually find that clause (13) is conflicting and we can construct the {\it conflict graph} in Figure~\ref{fig:imp-graph}, where every literal in the current decision level has incoming edges corresponding to its antecedents (except of course the decision literal). A special conflict node $\perp$ with incoming edges $\{\overline{x_{12}},x_{13}\}$ represents conflicting clause (13).

\tikzset{%
  CurNode/.style={draw, very thick, circle},
  DecNode/.style={draw, very thick, double, circle},
  PreNode/.style={draw, fill=cyan!20, circle},
OffPath/.style={draw, circle},
OnDL/.style={->, >= Latex},
PreDL/.style={->, >=Latex, dashed}}

\begin{figure}[t]
\begin{center}
\begin{tikzpicture}
[scale=0.09]
\node [DecNode] (x1) at (0,20) {$x_1$};
\node [CurNode] (x2) at (15,30) {$x_2$};
\node [CurNode] (x3) at (15,20) {$\overline{x_3}$};
\node [CurNode] (x4) at (15,10) {$x_4$};
\node [PreNode] (y2) at (15,-2) {$\overline{y_2}$};
\node [PreNode] (y1) at (30,40) {$\overline{y_1}$};
\node [CurNode] (x5) at (30,20) {$x_5$};
\node [PreNode] (y3) at (30,-2)  {$y_3$};
\node [CurNode] (x6) at (45,30) {$\overline{x_6}$};
\node [CurNode] (x7) at (45,20) {$x_7$};
\node [CurNode] (x9) at (45,10) {$\overline{x_9}$};
\node [PreNode] (y5) at (70,42) {$y_5$};
\node [CurNode] (x8) at (60,30) {$x_8$};
\node [PreNode] (y6) at (80,42) {$\overline{y_6}$};
\node [CurNode] (x11) at (75,24) {$x_{11}$};
\node [CurNode] (x10) at (75,12) {$\overline{x_{10}}$};
\node [PreNode] (y4) at (75,-2) {$y_4$};
\node [CurNode] (x12) at (90,28) {$\overline{x_{12}}$};
\node [CurNode] (x13) at (90,15) {$x_{13}$};
\node [CurNode] (c) at (105,20) {$\perp$};

\draw [blue, ultra thick] plot [smooth, tension=0.5] coordinates { (108, 8) (50, 0) (36, 3) (36,20) (42, 35) (70, 36) (90, 35.5) (108,  30)};

\draw[OnDL] (x1) -- (x2);
\draw[OnDL] (x1) -- (x3);
\draw[OnDL] (x1) -- (x4);
\draw[PreDL] (y1) -- (x2);
\draw[OnDL] (x2) -- (x5);
\draw[OnDL] (x3) -- (x5);
\draw[OnDL] (x4) -- (x5);
\draw[PreDL] (y2) -- (x4);
\draw[PreDL] (y1) -- (x6);
\draw[OnDL] (x5) -- (x6);
\draw[PreDL] (y3) -- (x5);
\draw[OnDL] (x6) -- (x8);
\draw[OnDL] (x5) -- (x7);
\draw[OnDL] (x5) -- (x9);
\draw[PreDL] (y3) -- (x9);
\draw[OnDL] (x7) -- (x8);
\draw[PreDL] (y4) -- (x9);
\draw[PreDL] (y4) -- (x10);
\draw[OnDL] (x9) -- (x10);
\draw[OnDL] (x9) -- (x11);
\draw[OnDL] (x10) -- (x13);
\draw[PreDL] (y6) -- (x11);
\draw[PreDL] (y5) -- (x11);
\draw[OnDL] (x8) -- (x11);
\draw[OnDL] (x11) -- (x12);
\draw[OnDL] (x11) -- (x13);
\draw[OnDL] (x12) -- (c);
\draw[OnDL] (x13) -- (c);
%
%\draw[PreDL] (n1) edge [bend right=13] (n4);
\end{tikzpicture}   
\end{center}
  \caption{Conflict graph associated with Example~\ref{ex:conflict}. If 1UIP learning is applied, we generate lemma $y_1 \lor \lnot y_3 \lor \lnot y_4 \lor \lnot y_5 \lor y_6 \lor \lnot x_5$. White nodes belong to the current decision level, whereas blue ones are from previous decision levels.}
  \label{fig:imp-graph}
\end{figure}

The graph clearly shows that if we set $x_1$ to true, together with the literals of previous decision levels (the $y$'s), we obtain a conflict. However, the same happens with $x_5$, since any path from $x_1$ to the conflict necessarily goes through $x_5$. Literals with this property are called {\it Unique Implication Points} (UIPs), of which we only have $x_1$ and~$x_5$. Since $x_5$ is the one closest to the conflict we call it {\it First Unique Implication Point} (1UIP)~\cite{ConflictAnalysis}. It is easy to see that if we set $x_5$ and the $y$ literals that enter the cut delimited by the blue line, unit propagation derives the same conflict. Hence, since they cannot be simultaneously true, we can learn 
$y_1 \lor \lnot y_3 \lor \lnot y_4 \lor \lnot y_5 \lor y_6 \lor \lnot x_5$. The quality of a lemma can be assessed by its {\it Literal Block Distance (LBD)}~\cite{LBDAudemardSimon}: the number of different decision levels of the literals in the lemma. The lower the LBD, the better the lemma. In our case,  if we are at decision level 5, $y_3$, $y_4$
    and $y_6$ belong to decision level 2, and $y_1$ and $y_5$ to decision level 4, the LBD of the lemma is 3.    
    \hfill $\Box$
    \end{exa}

\smallskip
\noindent
{\it Resolution.} Given two clauses $l \lor C$ and $\lnot l \lor D$, the {\it resolution} inference rule allows one to derive the logical consequence $C \lor D$. It is well known that the lemma derived in Example~\ref{ex:conflict} can be obtained via a series of resolution steps that start with the conflicting clause, and resolve with reasons of literals in the reverse order in which they were added to the assignment. In fact, it has been proved~\cite{CDCLResolutionKnot,CDCLResolutionAtserias} that resolution and CDCL (with restarts) are polynomially equivalent, and hence classes of formulas that are hard for resolution; e.g., the pigeonhole principle - PHP~\cite{haken1985intractability}, 
or Tseitin formulas~\cite{TseitinHardForResolution}) are also hard for CDCL SAT solvers.

\medskip
\noindent
{\it Extended Resolution (ER).} Given two literals $l_1$ and $l_2$, the {\it extended resolution}~\cite{tseitin1983complexity} rule allows us to introduce clauses representing the definition $z \leftrightarrow l_1 \lor l_2$. ER can be substantially more powerful than resolution; for instance, it allows polynomial size proofs of PHP~\cite{PigeonholeExtendedResolution}. Incorporating ER to CDCL solvers could potentially enable them to solve such formulas in polynomial time. 
However, we lack good methods to incorporate ER into CDCL proof search.
This is precisely the aim of this paper, namely, incorporating a restricted version of ER into CDCL.

%\albert{Is it known whether there are polynomial proofs in ER of Tseitin formulas? Which paper should we cite?}
%\sam{This is definitely known. It is not addressed explicitly there, but from my PHP paper, 1986?, it is clear that Tseitin formulas have
%poly size Frege proofs and hence poly size ER proofs. Also, the
%Buss-Thapen DRAT paper has explicit SPR- proofs for Tseitin 
%tautologies, and ER simulates SPR-.}

\section{Dual Implication Points}
\label{sec:DIP-detection}
As discussed in the previous section, UIPs
are crucial for conflict clause learning in the CDCL algorithm.
We now introduce a new concept of a \emph{Dual Implication Point} (DIP) that gives
a tool for analyzing the conflict graph. Its applications include
discovering new implied 2-clauses, introducing new variables by extension,
and learning clauses involving the extension variables.  The idea behind
a dual implication point is that it consists of a pair of vertices (literals)
in the conflict graph that disconnects or ``separates'' the decision literal from the contradiction.\footnote{Perhaps
``Dual Implication Pair'' would be a better name than ``Dual Implication Point', since
a DIP is a pair of literals.  We use ``Point'' however to match the terminology
of ``Unique Implication Points''.}  
More precisely, a DIP is
defined to be a pair $\{x,y\}$ of literals such that all paths
in the conflict graph to the vertex~$\perp$ pass through 
at least one of $x$ and~$y$ and such that neither $x$ nor~$y$ is
a UIP.
In contrast, UIP is a {\em single} literal
that separates the decision literal from the conflict.

We use the
example in Figure~\ref{fig:imp-graph} to illustrate 
the concept of DIPs and their potential
applications. Recall that $x_5$ is the first UIP. In our applications,
we are seeking DIPs between the first UIP and the conflict node~$\perp$.
An obvious DIP is the pair $\overline {x_{10}}$ and $x_{11}$, since it is immediately 
clear that any path from~$x_5$ (or from~$x_1$) must pass through one of
$\overline {x_{10}}$ or~$x_{11}$.  On the other hand, the pair 
$\overline {x_{10}}$ and~$x_8$ is not a DIP since there are paths from $x_1$ to~$\perp$
that avoid these two literals; namely, any path that includes the edge from
$\overline {x_9}$ to~$x_{11}$.
There are several other DIPs in Figure~\ref{fig:imp-graph}: a complete list is given in
Figure~\ref{fig:allDIPsAndLearned}. 

\begin{figure}[t]
\centering
\begin{tabular}{c|c|c|c}
Dual & \\
Implication & Extension & Post-DIP & Pre-DIP \\ 
Point(DIP) & Variable & Learned Clause & Learned Clause \\
\hline
$\overline {x_{12}} , x_{13}$ & $z \leftrightarrow (\overline {x_{12}} \land x_{13})$ &  $\lnot z$ 
         & $\lnot x_5 \lor y_1 \lor \lnot y_3 \lor \lnot y_4 \lor \lnot y_5 \lor y_6 \lor z$ \\
$x_{11} , x_{13}$ &  $z \leftrightarrow (x_{11} \land x_{13})$ & $\lnot z$ 
         & $\lnot x_5 \lor y_1 \lor \lnot y_3 \lor \lnot y_4 \lor \lnot y_5 \lor y_6 \lor z$ \\
$\overline {x_{10}} , x_{11}$ &  $z \leftrightarrow (\overline {x_{10}} \land x_{11})$ & $\lnot z$ 
         & $\lnot x_5 \lor y_1 \lor \lnot y_3 \lor \lnot y_4 \lor \lnot y_5 \lor y_6 \lor z$ \\
$\overline {x_9} , x_{11}$ &  $z \leftrightarrow (\overline {x_9} \land x_{11})$ & $\lnot z \lor \lnot y_4$ 
         & $\lnot x_5 \lor y_1 \lor \lnot y_3 \lor \lnot y_4 \lor \lnot y_5 \lor y_6 \lor z$ \\
$x_8, \overline {x_9}$ &  $z \leftrightarrow (x_8 \land \overline {x_9})$ & $\lnot z \lor \lnot y_4 \lor \lnot y_5 \lor y_6$
         & $\lnot x_5 \lor y_1 \lor \lnot y_3 \lor \lnot y_4 \lor z$ \\
\end{tabular}
\caption{The complete list of DIPs for the conflict graph of
Figure~\ref{fig:imp-graph}.  The DIP-learnable clauses involve
the new extension variable~$z$ that can be introduced for that
DIP. (The extension clauses defining~$z$ must also be learned; e.g., for the
first line, the extension clauses express that $z \leftrightarrow \overline{x_{12}} \land x_{13}$.)}
\label{fig:allDIPsAndLearned}
\end{figure}

Figure~\ref{fig:allDIPsAndLearned} also shows how a DIP pair can be used to
introduce a variable~$z$ via extension, and the associated clauses that
can be learned. For example, in the third line,  the new extension variable~$z$ is 
introduced with the three clauses $\lnot z \lor \lnot x_{10}$,
$\lnot z \lor x_{11}$ and $x_{10} \lor \lnot x_{11} \lor \lnot z$ which express
the condition $z \leftrightarrow (\lnot x_{10} \land x_{11})$. From the conflict graph,
this allows inferring the clauses $\lnot z$ (the ``post-DIP'' learned clause)
and $\lnot x_5 \lor y_1 \lor \lnot y_3 \lor \lnot y_4 \lor \lnot y_5 \lor y_6 \lor z$
(the ``pre-DIP'' learned clause), both of which will be formally defined
  in the next paragraph. Since the post-DIP learned clause does not
have any variables from lower decision levels, we can
also infer the 2-clause $x_{10} \lor \lnot x_{11}$ either instead of
or in addition to introducing~$z$ and the pre- and post-DIP clauses.  Introducing
2-clauses in this way might be helpful for CDCL solvers that do special processing
of 2-clauses; for instance, in the work of Bacchus~\cite{Bacchus:EnhancingDP} or the
recent work of Biere et al.~\cite{BFW:CadiBack} or Buss et al.~\cite{BKVW:DualDFS}. 

% In general, for any literals $a,b$ that form a DIP in the above fashion, 
% we may introduce an extension variable $z \leftrightarrow a \land b$, and learn
% a pre-DIP clause of the form $\lnot f \lor \lnot C \lor z$ and
% a post-DIP clause of the form $\lnot z \lor \lnot D$, where $f$~is
% the first UIP and where $C$ and~$D$ are conjunctions of literals
% that were set true at lower levels. 

In general, for any literals $a,b$ that form a DIP in the above
fashion, we may introduce an extension variable $z \leftrightarrow a
\land b$, and learn (1) a pre-DIP clause of the form $\lnot f \lor
\lnot C \lor z$ (i.e. $f \land C \rightarrow z$), where $f$~is the
first UIP and $C$ the set of literals from previous decision levels
that have an edge in the conflict graph to any literal appearing after
the first UIP and no later than the DIP pair; and (2) a post-DIP
clause of the form $\lnot z \lor \lnot D$ (i.e. $z \land D \rightarrow
\perp$), where $D$ contains those literals from previous decision
levels with an edge to any literal appearing strictly after the DIP
pair. The correctness of adding these two clauses is guaranteed
because they can be generated via a sequence of resolution steps on
the formula and the clauses that define the extended variable $z$.
We can also prove their correctness with a more
  conflict-graph oriented argument: it should be obvious from the
  conflict graph that if we assert the 1UIP $f$ and all literals in
  $C$, we can obtain $a$ and $b$ via unit propagation. This means that
  $f \land C \rightarrow a \land b$ is a valid formula which, after
  replacing $a\land b$ by their definition $z$, becomes the pre-DIP
  clause $\lnot f \lor \lnot C \lor z$. Similarly, if we assert $a,b$
  and all literals in $D$, we get a conflict by unit propagation. This
  means that $\lnot a \lor \lnot b \lor \lnot D$ is a valid clause. If we apply
  two resolution steps with the clauses $a \lor \lnot z$ and $b \lor \lnot z$
  we obtain the post-DIP clause $\lnot z \lor \lnot D$.

For example, the last line of the table in
Figure~\ref{fig:allDIPsAndLearned} shows the case $z \leftrightarrow
(x_8 \land \overline {x_9})$, where $f$ is~$x_5$, and $C$ and~$D$ are
$\lnot y_1 \land y_3 \land y_4$ and $y_4 \land y_5 \land \lnot y_6$,
respectively. In Section~\ref{sec:DIP-CDCL} we discuss many possible
ways that DIP extension variables and clauses may be introduced into
CDCL solvers.  The rest of this section describes how to find DIPs
and its content is not strictly necessary to understand the
  rest of the paper.

\subsection{An Algorithm for Finding DIPs}
\label{sec:algorithm}

\tikzset{%
  CurNode/.style={draw, very thick, circle},
  DIPNode/.style={draw,very thick, double, circle, fill=cyan!20},
  DecNode/.style={draw, very thick, double, circle},
OffPath/.style={draw, circle},
OnDL/.style={->, >= Latex},
PreDL/.style={->, >=Latex, dashed}}

\begin{figure}[t]
\begin{minipage}{7cm}
  \begin{tikzpicture}
[scale=0.08]
\node [CurNode] (x1) at (0,20) {$x_1$};
\node [CurNode] (x3) at (20,35) {$x_3$};
\node [CurNode] (x4) at (20,20) {$x_4$};
\node [DIPNode] (x2) at (20,5) {$x_2$};
\node [DIPNode] (x5) at (40,35) {$x_5$};
\node [CurNode] (x6) at (40,20) {$x_6$};
\node [CurNode] (x7) at (40,5) {$x_7$};
\node [CurNode] (c) at (60,20) {$\perp$};

\draw[OnDL] (x1) -- (x3);
\draw[OnDL] (x1) -- (x4);
\draw[OnDL] (x1) -- (x2);
\draw[OnDL] (x3) -- (x5);
\draw[OnDL] (x4) -- (x5);
\draw[OnDL] (x2) -- (x6);
\draw[OnDL] (x2) -- (x7);
\draw[OnDL] (x5) -- (c);
\draw[OnDL] (x6) -- (c);
\draw[OnDL] (x7) -- (c);
%
%\draw[PreDL] (n1) edge [bend right=13] (n4);
\end{tikzpicture}
\end{minipage}
\!\!\!\!\!\!\!\!\!\!\!\!
\begin{minipage}{3cm}
  \begin{tabular}{|c|l|}\hline
{\bf Trail} & {\bf Derived clause}  \\\hline
    $\perp$ & $\lnot x_5 \lor \lnot x_6 \lor \lnot x_7$\\\hline
    $x_5$ & $\lnot x_3 \lor \lnot x_4 \lor \lnot x_6 \lor \lnot x_7$\\\hline
    $x_7$ & $\lnot x_3 \lor \lnot x_4 \lor \lnot x_6 \lor \lnot x_2$\\\hline
    $x_6$ & $\lnot x_3 \lor \lnot x_4 \lor \lnot x_2$\\\hline
    $x_4$ & $\lnot x_3 \lor \lnot x_1 \lor \lnot x_2$\\\hline
    $x_3$ & $\lnot x_1 \lor \lnot x_2$\\\hline
    $x_2$ & $\lnot x_1$ \\\hline
    $x_1$ & \\\hline
  \end{tabular}

\end{minipage}

  \caption{Conflict graph associated with Example~\ref{ex:failed-naive-dip}, with the literals in blue being the only DIP. Next to it, the resulting trail and the sequence of intermediate clauses obtained if standard 1UIP learning is applied.}
  \label{fig:failed-naive-dip}
\end{figure}

A conventional CDCL algorithm maintains a trail of the literals set
  true, in the order they were set, and this allows finding the UIP very quickly.
  The idea is to initialize a clause $C$ as the conflicting clause and,
  following the order of the trail from the most recent literal $l$ backwards,
  apply
  resolution between $C$ and the reason of $l$ to get a new clause $C$. This is repeated
  until $C$ only has only one literal of the current decision level.

One could think that finding DIPs can be done by applying the same algorithm
  and stopping when $C$ has exactly two literals of the current decision level.
  This is what is done in~\cite{PD08Learning} but, as we now show, this procedure is not complete for detecting DIPs.
  \begin{exa}
    \label{ex:failed-naive-dip}
    Consider the set of clauses
  $\{
  \lnot x_1 \lor x_2,\allowbreak \;
  \lnot x_1 \lor x_3,\allowbreak \;
  \lnot x_1 \lor x_4,\allowbreak \;
  \lnot x_3 \lor \lnot x_4 \lor x_5,\allowbreak \;
  \lnot x_2 \lor x_6,\allowbreak \;
  \lnot x_2 \lor x_7,\allowbreak \;
  \lnot x_5 \lor \lnot x_6 \lor \lnot x_7
  \}$.
  Setting decision $x_1$ and applying unit propagation finds a
  conflict. This is illustrated in the conflict graph of
  Figure~\ref{fig:failed-naive-dip}, where the only DIP is the pair
  $\{x_2,x_5\}$. In the same figure, we can see the corresponding trail
  and the intermediate clauses obtained in a standard conflict
  analysis procedure. Note that the procedure is unable to detect the DIP.
  \end{exa}

  Finding the DIPs is much more complex than finding the UIP,
  and requires several traversals of the conflict graph between the
  first-UIP and the conflict node. Nonetheless, it is possible to find
  all DIPs very quickly, even in linear time.

We recast the DIP-finding
problem in terms of a general directed graph~$G$. Let $G = (V, E)$ be
a directed graph with two distinguished vertices $s$ and~$t$. We assume
that $s$ is 2-connected to~$t$ in that there is a pair of vertex-disjoint
paths $\pi_1$ and~$\pi_2$ from $s$ to~$t$ that share no vertices apart 
from $s$ and~$t$.  By the vertex-version of Menger's theorem~\cite{Menger:Kurventheorie}, 
either there are in fact {\em three} vertex-disjoint paths from $s$ to~$t$ or
there is at least one pair of vertices $\{a, b\}$ with neither $a$ nor~$b$
in~$\{s,t\}$ so that every path from $s$ to~$t$ passes through at least
one of $a$ or~$b$.  Such a pair $\{a,b\}$, if it exists, is called a
{\em Two Vertex Dominator} (TVD). 

Our goal is to find all possible TVDs efficiently and in linear
time. An algorithm for this is discussed in detail in our companion
paper~\cite{BGO:TVBs}; for space reasons, we give only an abbreviated
discussion of the algorithm here and direct the reader
  who wants a deep understanding of the algorithm to~\cite{BGO:TVBs}. The first step is to find two
vertex disjoint paths from $s$ to~$t$: this is done by greedily
finding a path from $s$ to~$t$ and then using an augmenting
path\footnote{This can also be computed in linear time by
  applying Ford-Fulkerson's algorithm to a small modification of the conflict
  graph in order look for a flow of 2.} construction to find two
vertex-disjoint paths from $s$ to~$t$.  An example is shown in
Figure~\ref{fig:DIPalgorithm}, where the two paths are $a_0,\ldots,
a_\ell$ and $b_0,\ldots,b_k$ where $a_0 = b_0 = s$ and $a_\ell = b_k =
t$. These paths are called $\pi_a$ and~$\pi_b$, respectively.
Henceforth, a {\em path} is a directed path without any repeated
nodes. The \emph{internal} vertices of a path~$\pi$ are the vertices
on~$\pi$ other than the first and last vertices. Two paths are said to
be \emph{vertex-disjoint} if they have no internal vertices in common.
A path~$\pi$ \emph{avoids $\pi_a$ and~$\pi_b$} if it is
vertex-disjoint from both paths.

\begin{figure}[t]
\centering
\begin{tikzpicture}[scale=0.14,
every node/.style={draw, fill, circle, inner sep=0, minimum size=4pt}, 
OnPath/.style={->, >=Latex, thick},
OffPath/.style={->, >=Latex},
every label/.append style={rectangle, inner sep = 2pt} ]
\node (s) at (0,0) [label=left:$s$] {};
\node (a1) at (10,5) [label=above:{$a_1{=}a^\prime_1$}] {};
\node (a2) at (20,5) [label=above:{$a_2{=}a^\prime_2$}] {};
\node (a3) at (30,5) [label=above:{$a_3{=}a^\prime_3$}] {};
\node (a4) at (40,5) [label=above:{$a_4$}] {};
\node (a5) at (50,5) [label=above:{$a_5{=}a^\prime_4$}] {};
\node (b1) at (10,-5) [label=below:$b_1{=}b'_1$] {};
\node (b2) at (20,-5) [label=below:$b_2{=}b'_2$] {};
\node (b3) at (30,-5) [label=below:$b_3$] {};
\node (b4) at (40,-5) [label=below:$b_4$] {};
\node (b5) at (50,-5) [label=below:$b_5{=}b'_3$] {};
\node (t) at (60,0) [label=right:$t$] {};
\node (c) at (26,-2) [label=above:$c$] {};
\node (d) at (33,-0) [label=below:$d$] {};
\draw [OnPath] (s) -- (a1);
\draw [OnPath] (a1) -- (a2);
\draw [OnPath] (a2) -- (a3);
\draw [OnPath] (a3) -- (a4);
\draw [OnPath] (a4) -- (a5);
\draw [OnPath] (a5) -- (t);
\draw [OnPath] (s) -- (b1);
\draw [OnPath] (b1) -- (b2);
\draw [OnPath] (b2) -- (b3);
\draw [OnPath] (b3) -- (b4);
\draw [OnPath] (b4) -- (b5);
\draw [OnPath] (b5) -- (t);
\draw [OffPath] (a3) edge[bend right] (a5); 
\draw [OffPath] (b2) -- (c);
\draw [OffPath] (c) -- (d);
\draw [OffPath] (d) -- (a4);
\draw [OffPath] (d) edge[bend left=12] (b5);
\end{tikzpicture}
\caption{Nodes $a_4$, $b_3$ and~$b_4$ are bypassed and thus not in any TVD.
The path $b_2, c, d, a_4$ is a crossing separator that prevents
any of $a_1$, $a_2$ and~$a_3$ from being paired with~$b_5$ 
to form a TVD pair.  All other
pairs $\{a'_i,b'_j\}$ are valid TVD pairs.}
\label{fig:DIPalgorithm}
\end{figure}

Once the two vertex-disjoint paths are fixed, we
can state the following definitions and theorems:

\begin{defiC}[\rm \cite{BGO:TVBs}]
A node~$a_i$ on~$\pi_a$ is \emph{bypassed} 
if there are $j<i<j'$ and a path~$\pi$ from $a_j$ to~$a_{j'}$
such that $\pi$~avoids $\pi_a$ and~$\pi_b$.
A node~$b_i$ being bypassed is defined similarly.
\end{defiC}

\begin{defi}
Two nodes $a_i$ and~$b_j$ have a \emph{crossing separator}
if there are nodes $a_{i'}$ and~$b_{j'}$ joined by a path~$\pi$ that avoids both $\pi_a$ and~$\pi_b$ such
that either (a)~$i'<i$ and $j'>j$ and $\pi$~is a path
from $a_{i'}$ to~$b_{j'}$, or (b)~$i'>i$ and $j'<j$
and $\pi$~is a path from $b_{j'}$ to~$a_{i'}$.
\end{defi}

\begin{thmC}[\rm \cite{BGO:TVBs}]\label{thm:TVDproperty}
For $0<i<\ell$ and $0<j<k$, the two nodes $a_i$ and $b_j$
form a two-vertex dominator (TVD) if and only if
$a_i$ and~$b_j$ do not have a crossing separator and 
neither $a_i$ nor~$b_j$ is bypassed.
\end{thmC}

Theorem~\ref{thm:TVDproperty} is proved in~\cite{BGO:TVBs}.
The theorem holds for both
acyclic and cyclic directed graphs; however, for our applications to
CDCL we are interested only in acyclic graphs since the conflict
graph is always acyclic.

A consequence of Theorem~\ref{thm:TVDproperty} is that
the set of all TVDs can be compactly represented in linear size,
even though there can be quadratically many TVDs. Let 
$a'_1, \ldots a'_{\ell'}$ be the subsequence of 
the internal nodes $a_1.\ldots,a_{\ell-1}$
of path~$\pi_a$ that, according to the conditions
of Theorem~\ref{thm:TVDproperty}, are in at least one TVD pair.
Let $b'_1,\ldots,b'_{k'}$ be the corresponding subsequence
of the internal nodes of~$\pi_b$. Then, for each~$a'_i$
there are $m \le n$ such that $a'_i$ forms a TVD
pair with each $b'_j$ with $m\le j \le n$.
Dually, for each~$b'_i$
there are $m \le n$ such that $b'_i$ forms a TVD
pair with each $a'_j$ with $m\le j \le n$.
%(See Figure~\ref{thm:TVDproperty}.)

\begin{exa}
In the conflict graph of Figure~\ref{fig:imp-graph},
consider the portion of the graph between the first UIP $x_5$
and the contradiction~$\perp$.  We can take path~$\pi_a$
to be $x_5, x_7, x_8, x_{11}, \overline{x_{12}},\perp$ and path~$\pi_b$ to
be $x_5, \overline{x_9}, \overline{x_{10}}, x_{13}, \perp$.
The node $x_7$ is bypassed by the path $x_5, \overline{x_6}, x_8$, so it
cannot be part of a TVD. There are two crossing separator paths:
the first is the edge from $\overline{x_9}$ to~$x_{11}$; the second
is the edge from $x_{11}$ to~$x_{13}$.
Therefore,  the possible TVD pairs can be described in a table as:
\begin{center}
\begin{tabular}{c|c||c|c}
Node & forms a TVD & Node & forms a TVD \\
on $\pi_a$ & pair with & on $\pi_b$ & pair with \\
\hline
$x_8$ & $\overline {x_9}$ & $\overline{x_9}$ & $x_8, x_{11}$ \\
$x_{11}$ & $\overline{x_9}, \overline x_{{10}}, x_{13}$ & $\overline{x_{10}}$ & $x_{11}$ \\
$\overline{x_{12}}$ & $x_{13}$ & $x_{13}$ & $x_{11}, \overline x_{12}$  
\end{tabular}
\end{center}
In this example, one node, $x_7$, was bypassed. It is 
also possible that non-bypassed nodes are eliminated
just by the crossing separators. For example, if there were an
additional edge from $x_7$ to~$\overline{x_{10}}$, then
the crossing separator condition would imply that 
$x_8$ and~$x_{11}$ are not part of any TVD pair. In this
case, $x_8$ and~$x_{11}$ would not be included among
the $a'_i$ nodes.    
\end{exa} 

Theorem~\ref{thm:TVDproperty} is used in~\cite{BGO:TVBs} to give
an efficient, linear time algorithm for finding all
TVDs. The algorithm has five phases. 
The first two phases find two vertex-disjoint paths
from $s$ to~$t$; the next phase scans the graph from $t$ to~$s$
to discover all relevant paths that avoid $\pi_a$ and~$\pi_b$;
the fourth phase uses this to discard bypassed nodes and collect
information on crossing separators; finally the fifth phase
computes the compressed representation of all possible TVDs.
Full details are given~\cite{BGO:TVBs}, which is available online
in preprint form. In our experiments with the implication graphs constructed by the underlying CDCL solver, the time overhead in finding the TVDs is negligible.

\section{Extension Variables from Dual Implication Points}\label{sec:}

This section discusses how DIPs can be used for introducing extension
variables and implement the ERCL method for learning ER clauses in CDCL solver. We first present an example.

\subsection{An example with a grid Tseitin principle}\label{sec:TseitinExample}

% We now present a small example of how DIP clause learning works with
% a Tseitin principle on a $3\times 3$-grid. Figure~\ref{fig:TseitinExample}
% shows the grid and the associated clauses, expressed as XOR conditions. 
% For a CDCL refutation, the XOR conditions are replaced with 
% clauses. Only one vertex has charge~1, so the clauses are unsatisfiable.

Given a graph where every vertex has a {\it charge}, a number which is
$0$ or $1$, a Tseitin formula~\cite{tseitin1983complexity} is created
by considering one variable per edge, and adding one constraint per
vertex $v$ expressing that the sum of the variables of all edges
incident to $v$ modulo 2 is equal to its charge. The CNF version that
we consider converts each {\it xor} constraint into clauses by simple
enumeration of falsifying assignments. It is easy to see that the formula is unsatisfiable if
and only if the sum of all charges is odd. Here we consider as a graph
the $3\times 3$-grid depicted in Figure~\ref{fig:TseitinExample} and
only one vertex has charge~1, so the clauses are unsatisfiable.

\begin{figure}[t]
\begin{center}
\begin{minipage}{2cm}
\begin{tikzpicture}[scale=0.12,
Gnode/.style={draw, circle, inner sep = 2pt}]
\node[Gnode] (a) at (0,0) {$\scriptsize 1$};
\node[Gnode] (b) at (10,0) {$0$};
\node[Gnode] (c) at (20,0) {$0$};
\node[Gnode] (d) at (0,-10) {$0$};
\node[Gnode] (e) at (10,-10) {$0$};
\node[Gnode] (f) at (20,-10) {$0$};
\node[Gnode] (g) at (0,-20) {$0$};
\node[Gnode] (h) at (10,-20) {$0$};
\node[Gnode] (i) at (20,-20) {$0$};
\draw[-] (a) edge node[above]{$e_1$} (b);
\draw[-] (b) edge node[above]{$e_2$} (c);
\draw[-] (a) edge node[left]{$e_3$} (d);
\draw[-] (b) edge node[right]{$e_4$} (e);
\draw[-] (c) edge node[right]{$e_5$} (f);
\draw[-] (d) edge node[above]{$e_6$} (e);
\draw[-] (e) edge node[above]{$e_7$} (f);
\draw[-] (d) edge node[left]{$e_8$} (g);
\draw[-] (e) edge node[right]{$e_9$} (h);
\draw[-] (f) edge node[right]{$e_{10}$} (i);
\draw[-] (g) edge node[below]{$e_{11}$} (h);
\draw[-] (h) edge node[below]{$e_{12}$} (i);
\end{tikzpicture}
\end{minipage}
\hspace*{7em}
\begin{minipage}{5.5cm}
\begin{tabular}{r@{\,$=$\,}c@{\hspace*{20pt}}r@{\,$=$\,}c}
$e_1 \oplus e_3$ & 1 &
    $e_5 \oplus e_7 \oplus e_{10}$ & 0 \\
$e_1 \oplus e_2 \oplus e_4$ & 0  &
    $e_8 \oplus e_{11}$ & 0 \\
$e_2 \oplus e_5$ & 0 & 
    $e_9 \oplus e_{11} \oplus e_{12}$ & 0 \\
$e_3 \oplus e_6 \oplus e_8$ & 0 & 
    $e_{10} \oplus e_{12}$ & 0 \\
$e_4 \oplus e_6 \oplus e_7 \oplus e_9$ & 0 
\end{tabular}
\end{minipage}
\end{center}
\caption{An instance of the $3\times 3$-grid Tseitin principle. The nodes
are assigned a polarity in~$\{0,1\}$. The edges are labeled with variables.}
\label{fig:TseitinExample}
\end{figure}

The first steps of the CDCL solver are as follows: First $e_1$~is set true
as a decision literal, and $\overline {e_3}$ is (unit) propagated.
Second, $e_2$~is set true as a decision literal, and $\overline{e_4}$ and~$e_5$ are
propagated. Third, $e_6$~is set true as a decision literal, and $e_8$ and~$e_{11}$
are propagated. Fourth, $e_7$~is set true as a decision literal, and
$\overline{e_9}$, $\overline{e_{10}}$, and~$e_{12}$ are propagated. This gives
a contradiction, since the clause $e_{10}\lor \overline{e_{12}}$
(one of the two clauses from $e_{10} \oplus e_{12}= 0$) is
falsified.  Figure~\ref{fig:TseitinConflict}(a) shows the
complete conflict graph at this point. 

Examining the conflict graph at decision level~4, 
the first UIP is~$e_7$ and there are two DIPs available,
$\{\overline{e_9},\overline{e_{10}}\}$ and $\{\overline{e_{10}}, e_{12}\}$.
Selecting the former DIP, we introduce
a new variable~$x$ by extension 
as $x \leftrightarrow \overline{e_9} \land \overline{e_{10}}$. We can learn the 
additional pre- and post-DIP clauses: $e_4 \lor \overline{e_5} \lor \overline{e_6} \lor \overline{e_7} \lor x$
and $\overline x \lor \overline{e_{11}}$.

%\[
%e_4 \lor \overline{e_5} \lor \overline{e_6} \lor \overline{e_7} \lor x .
% \qquad \hbox{and} \qquad
% \overline x \lor \overline{e_{11}}
%\]
We next backtrack to decision level~3, unsetting $e_7$, $e_9$, $e_{10}$ and~$e_{12}$.
Unit propagation at decision level~3 sets the new literal~$x$ and
the first UIP~$\overline{e_7}$ false and then sets literals $e_9$,
$e_{10}$, and~$\overline{e_{12}}$ true.  This yields a
contradiction with the clause $\overline{e_{10}} \lor e_{12}$.
In the conflict graph at decision level~3, 
the first UIP is~$e_6$ and there are two DIPs available,
namely 
$\{\overline{e_7}, \overline{e_{12}}\}$ and
$\{e_{10}, \overline{e_{12}}\}$.
If we select the first one, then we introduce a 
new literal $y$ defined by extension as 
$y \leftrightarrow \overline{e_7} \land \overline{e_{12}}$ and
can in addition learn the clauses $e_3 \lor e_4 \lor \overline{e_5} \lor \overline{e_6} \lor y$
and  $\overline{e_5} \lor \overline y$.
%\[
%e_3 \lor e_4 \lor \overline{e_5} \lor \overline{e_6} \lor y.
%\qquad \hbox{and} \qquad
%\overline{e_5} \lor \overline y 
%\]

We do not carry this example further, but note that
our experiments show that Tseitin tautologies (not
just on grid graphs) are examples where our experiments
show the DIP clause learning method is especially effective.
\begin{figure}[t]
\begin{center}
\begin{tikzpicture}[scale=0.14,
every node/.style={rectangle,inner sep = 2pt},
Dec/.append style={draw}, 
Imp/.style={->, >=Latex},
TopLvl/.style={->, >=Latex,thick}]
%Levels 1 and 2
\node[Dec] (e1) at (0,0) [label={[yshift=2pt]left:Level 1:}] {$e_1$};
\node (e3bar) at (13,0) {$\overline{e_3}$};
\node[Dec] (e2) at (0,-5) [label={[yshift=2pt]left:Level 2:}] {$e_2$};
\node (e4bar) at (10,-5) {$\overline{e_4}$};
\node (e5) at (20,-5) {$e_5$};
% Levels 3 and 4
\node[Dec] (e6) at (0,-11) [label={[yshift=2pt]left:Level 3:}] {$e_6$};
\node (e8) at (10,-11) {$e_8$};
\node (e11) at (22,-11) {$e_{11}$};
\node[Dec,thick] (e7) at (0,-18) [label={[yshift=2pt]left:Level 4:}] {{\boldmath $e_7$}};
\node (e9bar) at (12,-18) {{\boldmath $\overline{e_9}$}};
\node (e12) at (25,-18) {{\boldmath $e_{12}$}};
\node (e10bar) at (15,-23) {{\boldmath $\overline{e_{10}}$}};
\node (perp) at (30,-23) {{\boldmath $\perp$}};
% Levels 1 and 2, edges:
\draw[Imp] (e1) -- (e3bar);
\draw[Imp] (e1) -- (e4bar);
\draw[Imp] (e2) -- (e4bar);
\draw[Imp] (e2) edge[bend right=17] (e5);
% Levels 3 and 4, edges.
\draw[Imp] (e6) -- (e8);
\draw[Imp] (e3bar) edge[bend left=20] (e8.north);
\draw[Imp] (e8) -- (e11);
\draw[Imp] (e6) -- (e9bar);
\draw[TopLvl] (e7) -- (e9bar);
\draw[TopLvl] (e9bar) -- (e12);
\draw[Imp] (e11) -- (e12);
\draw[Imp] (e4bar) edge[bend left=20] (e9bar);
\draw[TopLvl] (e7) -- (e10bar);
\draw[Imp] (e5) edge[bend right=15] (e10bar);
\draw[TopLvl] (e12) -- (perp);
\draw[TopLvl] (e10bar)-- (perp);
% Label
\node at (-15,-8) {(a):};
\end{tikzpicture}

% Set x and y to 0.5cm instead of 1.0 cm to avoid "Dimension too large" error.
%    I halved x and y and doubled the scale from 0.14 to 0.28. (!)
\begin{tikzpicture}[x=0.5cm,y=0.5cm,scale=0.28,
every node/.style={rectangle,inner sep = 2pt},
Dec/.append style={draw}, 
Imp/.style={->, >=Latex},
TopLvl/.style={->, >=Latex,thick}]
%Levels 1 and 2
\node[Dec] (e1) at (0,0) [label={[yshift=2pt]left:Level 1:}] {$e_1$};
\node (e3bar) at (13,0) {$\overline{e_3}$};
\node[Dec] (e2) at (0,-5) [label={[yshift=2pt]left:Level 2:}] {$e_2$};
\node (e4bar) at (10,-5) {$\overline{e_4}$};
\node (e5) at (20,-5) {$e_5$};
% Level 3
\node[Dec] (e6) at (0,-11) [label={[yshift=2pt]left:Level 3:}] {{\boldmath $e_6$}};
\node (e8) at (12,-11) {{\boldmath $e_8$}};
\node (e11) at (24,-11) {{\boldmath $e_{11}$}};
\node (xbar) at (33,-11) {{\boldmath $\overline{x}$}};
\node (e7bar) at (42,-11) {{\boldmath $\overline{e_7}$}};
\node (e10) at (52,-11) {{\boldmath $e_{10}$}};
\node (e9) at (38,-18) {{\boldmath $e_9$}};
\node (e12bar) at (54,-18) {{\boldmath $\overline{e_{12}}$}};
\node (perp) at (64,-18) {{\boldmath $\perp$}};
% Levels 1 and 2, edges:
\draw[Imp] (e1) -- (e3bar);
\draw[Imp] (e1) -- (e4bar);
\draw[Imp] (e2) -- (e4bar);
\draw[Imp] (e2) edge[bend right=17] (e5);
% Level 3, edges.
\draw[TopLvl] (e6) -- (e8);
\draw[Imp] (e3bar) edge[bend left=20] (e8.north);
\draw[TopLvl] (e8) -- (e11);
\draw[TopLvl] (e11) -- (xbar);
\draw[TopLvl] (xbar) -- (e7bar);
\draw[Imp] (e4bar.340) to[out=-30, in=150] (e7bar.150);
\draw[Imp] (e5.-15) to[out=0] (e7bar.130);
\draw[Imp] (e5.15) to[out=0] (e10.140);
%\node[above right = -11pt and -14pt of e7bar] (e7barX) {};
%\draw[TopLvl] (e6) edge[bend right=15] (e7barX);
\draw[TopLvl] (e6) edge[bend right=15] (e7bar.200);
\draw[Imp] (e4bar) edge[bend right=15] (e9);
\draw[TopLvl] (e7bar) -- (e10);
\draw[TopLvl] (e6.330) to[out=-30, in=180] (e9.200);
%\draw[TopLvl] (e6.330) edge[bend right=8] (e9.200);
\draw[TopLvl] (e7bar) -- (e9);
\draw[TopLvl] (e9) -- (e12bar);
\draw[TopLvl] (e11) to[out=-20, in=155] (e12bar);
\draw[TopLvl] (e12bar) -- (perp);
\draw[TopLvl] (e10) -- (perp);
% Label
\node at (-15,-6) {(b):};
\end{tikzpicture}
\end{center}
\caption{The complete conflict graph from the example; (a)~at the first conflict and
(b)~at the second conflict.  Boxed literals are decision literals. Implications
at the top decision level are in bold.}
\label{fig:TseitinConflict}
\end{figure}

It is interesting to relate the Figure~\ref{fig:TseitinConflict}(b)
to Theorem~\ref{thm:TVDproperty} on DIPs.
In this example, there is only one way to choose the two vertex-disjoint paths. 
Namely, to let $\pi_a$ and~$\pi_b$ be the paths 
$e_6, e_8, e_{11}, \overline x, \overline{e_7}, e_{10}, \perp$
and $e_6, e_9, \overline{e_{12}}, \perp$. The
edge from $e_6$ to~$\overline{e_7}$ bypasses 
$e_8$, $e_{11}$ and~$\overline x$; 
and the edge from $e_{11}$ to~$\overline{e_{12}}$ is a crossing separator.
(The edge from $\overline{e_7}$ to~$e_9$ is a ``vacuous'' crossing separator
that does not actually remove any possible DIP pairs.)

It should be evident from this example that many conflicts have DIPs; indeed our experiments
reported below show that approximately
2/3 of the conflicts have at least one DIP and, 
very frequently there are quite a few choices
for DIPs.

\subsection{Extending CDCL with Dual Implication Points}
\label{sec:DIP-CDCL}\label{sec:xMapleLCM}

The use of DIPs in conflict analysis opens a large spectrum of possibilities. This section discusses some of them, with particular attention to the techniques that we have implemented.
Our present implementation, {\xmaple}, is a flexible framework that allows one to implement extended-resolution based techniques in a simple way.
It offers a set of clearly-specified functions that facilitate determining which extension variables to add, performing the corresponding addition, the possible posterior deletion or replacing definitions in clauses. Additionally, a set of heuristic choices to control when and how these steps are performed are provided, and replacing them by custom ones is a smooth task.
We want to remark that there are many more possible strategies for using DIPs than could be discussed. We believe that in this paper we merely scratched the surface of heuristics for exploiting DIPs, which is an indication of the potential of this approach.

\smallskip
\noindent
{\bf Choice of DIP.}
As mentioned, there is possibly a quadratic number of DIPs. Even though we could learn multiple DIPs at every conflict, with their corresponding lemmas, we decided to choose only one. The first possibility
we considered is to learn the DIP that is closest to the conflict, as we do with UIPs. This 
may often create a short post-DIP conflict but a long pre-DIP clause. Therefore, we considered the
possibility of choosing a DIP that splits the conflict graph into two
balanced regions. Ideally, that would result in two equally 
long pre and post-DIP clauses. Finally, we also implemented 
choosing a random DIP, to check whether any of the other two schemes
could outperform a random strategy. These heuristics are referred to as 
{\bf closest}, {\bf middle}, and {\bf random}, respectively.

\smallskip
\noindent
{\bf Filtering out bad-quality DIPs.} Learning a DIP whenever we find one would be too prolific and overwhelm the solver. In order to determine whether the DIP chosen
in the previous step had to be discarded, the first possibility we explored was again
inspired by 1-UIP learning, where learning glue clauses, i.e. having LBD equal to~$2$, is a desired situation. The first filtering mechanism we
implemented discarded all DIPs that did not have a glue post-DIP clause. Another possibility we considered is to wait for a DIP to occur a certain number of times before using it in DIP-based learning. In our implementation, we tried different numbers as the minimum threshold to introduce a DIP. A third possibility
is to use the activity-based heuristic of the literals in the DIP
to assess its quality: DIPs whose literals have high decision-heuristic scores
should be prioritized. In our implementation, we check whether the activity level
of the current DIP is higher than the
average activity level of the 20 most recently encountered DIPs; if so, the current DIP is
a candidate for DIP-learning, otherwise it is discarded.
We refer to these various techniques as {\bf glue}, {\bf occ} and {\bf act}, respectively.

\smallskip
\noindent
%\subparagraph*{Learning pre-DIP and post-DIP clauses.}
{\bf Learning pre-DIP and post-DIP clauses.}
In our implementation, we only considered two variants: one 
that always learns both the pre- and the post-DIP clauses ({\bf 2-clause}) 
and one
that only learns the post-DIP clause ({\bf 1-clause}).

\smallskip
\noindent
%\subparagraph*{Backjumping and asserting clauses.}
{\bf Backjumping and asserting clauses.}
Recall that when a new DIP extension variable~$z$ is introduced, it is
possible to learn
a pre-DIP clause of the form $\lnot f \lor \lnot C \lor z$ and
a post-DIP clause of the form $\lnot z \lor \lnot D$, where $f$~is
the first UIP and where $C$ and~$D$ are conjunctions of literals
that were set true at lower levels. (Note that $C$ and~$D$ may have 
literals in common.) Letting $\ell_C$ and $\ell_D$
be the maximum of the levels at which literals in $C$ and~$D$ (respectively)
were set, our implementation always backtracks to level~$\ell_D$. This makes
the post-DIP clause asserting, so $\lnot z$ is set at decision level~$\ell_D$.
Furthermore, if $\ell_C \le \ell_D$ and the pre-DIP clause is learned, then 
it is asserting and $\lnot f$ is set by
unit propagation at decision level~$\ell_D$, as it should be.

It would make no sense to backtrack to level $\ell_C$ when $\ell_C < \ell_D$
since then neither $\lnot z$ nor~$\lnot f$ would be propagated.
However, another possible strategy would be to backtrack to the
maximum of the decision levels $\ell_C$ and~$\ell_D$. 
This would mean $\lnot z$ and~$\lnot f$ are both 
propagated. The disadvantage of this when $\ell_C > \ell_D$ is that
it would mean $\lnot z$ is asserted by the post-DIP clause 
at level~$\ell_C$, whereas it could have been propagated at the previous decision
level~$\ell_D$.  This breaks a usual invariant
for CDCL solvers. This would only be possible in a solver that permits chronological 
backtracking~\cite{NadelRychin:chronological,MohleBiere:backtracking}; 
{\xmaple}, however, does not support this.

The previous reasoning needs some clarification for the case when the extension variable $z$ with definition $z \leftrightarrow l_1 \land l_2$ we want to use has already been introduced.
If $z$ is undefined or defined at the current decision level nothing changes.
We know it cannot be true at some previous level, because otherwise $l_1$ and $l_2$ would have been propagated at that level, and not in the current one as all literals belonging to a DIP do. If it is false at some previous level, then we have no guarantee that
the pre or the post-DIP clause is asserting at any decision level. Fortunately, that rarely happens in practice. However, we can always perform standard 1UIP learning or try to apply DIP-based conflict analysis starting with the clause
$z\lor \lnot l_1 \lor \lnot l_2$ that we can guarantee is conflicting at the current decision level.

%\sam{Albert mentioned that we need to account for the case where
%the DIP is asserted at some previous DL. Does the current discussion cover the intended point?}\albert{Fixed: added a paragraph. Might be removed if too technical}

\smallskip
\noindent
{\bf Replacing literals by extended variables.}
%\subparagraph*{Replacing literals by extended variables.}
In {\xmaple},
    every time a new lemma is learned, we try to replace some of its
    literals by extended variables. An extended variable $z \leftrightarrow l_1 \lor l_2$
    allows one to replace a lemma of the form $l_1 \lor l_2 \lor C$ by
    $z \lor C$. This is done by checking, for all pairs of literals in the lemma
    that appear in some extended variable definition whether they are part
    of the same definition. Lemmas that are too long or have large LBD are
    discarded to mitigate the cost of this operation.

%    There is a subtle point worth being mentioned. 

    If one wants to introduce literal $\lnot z$ and obtain some
    reduction in formula size, the lemma should be of the form $\lnot
    l_1 \lor C$ and there should be another clause of the form $\lnot
    l_2 \lor C$. In this case, they can be replaced by a single clause
    $\lnot z \lor C$. However, this situation can be expensive to
    detect since one has to traverse the whole database looking for a
    certain clause, and this operation should be repeated for every
    literal in the lemma. For this reason, {\xmaple} does not implement
    this. 

\smallskip
\noindent
%\subparagraph*{Deleting extended variables.} 
{\bf Deleting extended variables.}
As it happens with lemma
    learning, where keeping too many lemmas slows down unit
    propagation, managing too many extended variables might also be
    counterproductive.  Again following the analogy with learned
    lemmas, which are useful at some point of the search but might
    become inactive after a while, it is natural to think that
    extended variables follow the same behavior. All in all,
    it seems mandatory to consider the deletion of extended variables,
    which amounts to deletion all lemmas where they appear.

    Deletion of variables in {\xmaple} is scheduled to be performed
    every $1000$ conflicts. At that point, several strategies are
    possible: delete all variables, delete the ones with a minimum
    decision-heuristic activity, or delete the worst $k\%$ variables
    according to some criterion (e.g. their decision-heuristic
    activity). In our implementation, we do the latter with $k=50$.
    Note that variables appearing in the right-hand side of an
    extended-variable definition cannot be deleted. This is addressed
    by maintaining a counter for every variable that corresponds to the
    number of definitions where it is involved.

\section{Experimental Evaluation}
\label{sec:experiments}
We have implemented the DIP-based clause learning schemes described in
Section~\ref{sec:DIP-CDCL} on top of the {\xmaple} ER
framework\footnote{All sources used for this evaluation can be found
in \url{https://github.com/chjon/xMapleSAT/tree/main}.}. We started
our experimental evaluation running a variant of DIP-based learning on
benchmarks of the SAT Competition~\cite{SATcomp2023} and
comparing it to MapleLCM~\cite{MapleLCM}, the CDCL SAT solver on which
it is based. Even though there did not seem to be a systematic
improvement on all benchmarks, a few families with important speedups
were identified. We start this section by analyzing the impact of
DIP-based learning on these families and then move to final
considerations about the performance on the overall SAT
competition benchmarks.

For each benchmark family, we start by describing the problem they encode.
After that, we evaluate the impact of the different techniques
explained in Section~\ref{sec:DIP-CDCL}. In
particular, we first consider as a {\bf baseline} a version that
(i) finds DIPs in the {\bf middle} of the conflict graph, (ii) only adds a
DIP if it has occurred at least 20 times and (iii) always learns both the
pre- and post-DIP learning clauses. 
Different variants are obtained by changing only one of the three
previous design decisions at a time.  Regarding the type of DIP used,
we analyze the performance of {\bf closest}, {\bf random}, 
 and {\bf heuristic} , that is, the systems whose only
difference w.r.t. {\bf baseline} is that the type of DIP is changed.
Regarding the criterion used to discard a DIP, we implemented the {\bf
  glue} and the {\bf act} configurations, but due to their poor
performance we do not include these results. We present instead
results about for the {\bf occ5} and {\bf occ50} configurations, which
discard any DIP that has not occurred at least 5 or 50 times,
respectively. Finally, we evaluated the performance of the {\bf
  1-clause} variant only learns the post-DIP clause.

After that, we report on the performance of a variety of
state-of-the-art solvers, each with some distinguished characteristic:
{\kissat} 4.0.2~\cite{kissat} (an extremely efficient CDCL solver),
{\crypto} 5.12.1~\cite{cryptominisat} (support for XOR reasoning),
{\sbva}~\cite{SBVA} (introduction of new variables via SBVA and
winner of the main track of the 2023 SAT Competition),
{\glucoser}~\cite{Audemard2010ARO} (extended-resolution based CDCL
solver) and the best available configuration for {\xmaple}.

\subsection{Tseitin Formulas}

These formulas have already been described in Section~\ref{sec:TseitinExample}.
In this section, we consider
three types of unsatisfiable formulas generated by CNFgen~\cite{CNFGen}: (i) Tseitin
formulas on the grid, where the graph
is a rectangular grid and each vertex is connected to its 4
neighbors, except for the vertices on the boundaries of the grid which have
fewer neighbors, (ii) 4-regular Tseitin formulas, where the graph is a $4$-regular random one,
and (iii) 6-regular Tseitin formulas built on top of $6$-regular random graphs.
The time limit was set to 600 seconds per instance.

\begin{figure}[t]
  \centering
      {\includegraphics[width=0.32\linewidth]{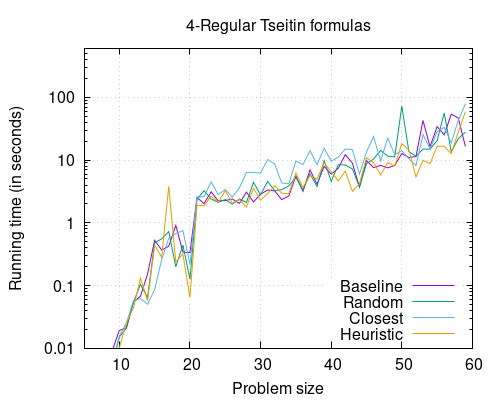}}
      {\includegraphics[width=0.32\linewidth]{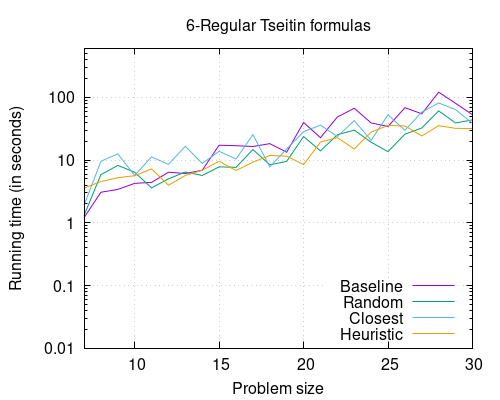}}
      {\includegraphics[width=0.32\linewidth]{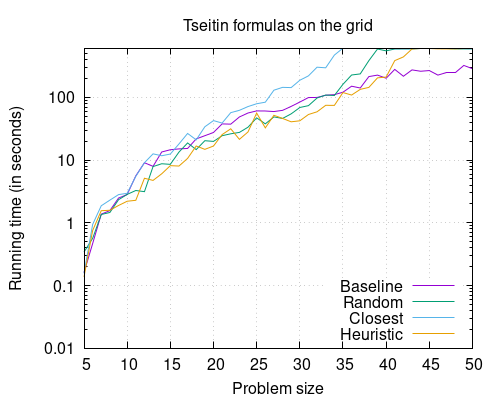}}\\\bigskip\bigskip
      {\includegraphics[width=0.32\linewidth]{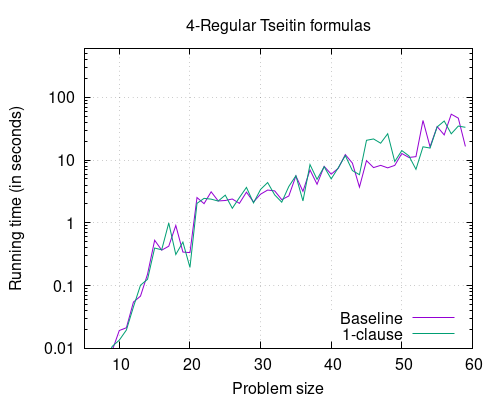}}
      {\includegraphics[width=0.32\linewidth]{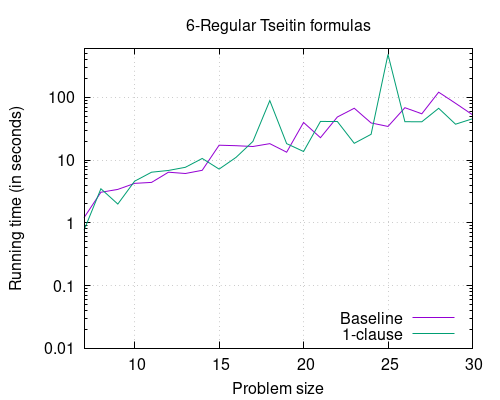}}
      {\includegraphics[width=0.32\linewidth]{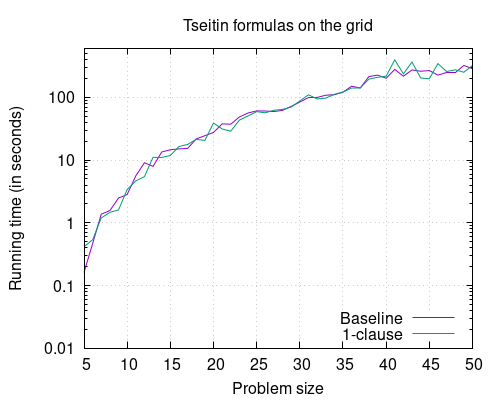}}\\\bigskip
      {\includegraphics[width=0.32\linewidth]{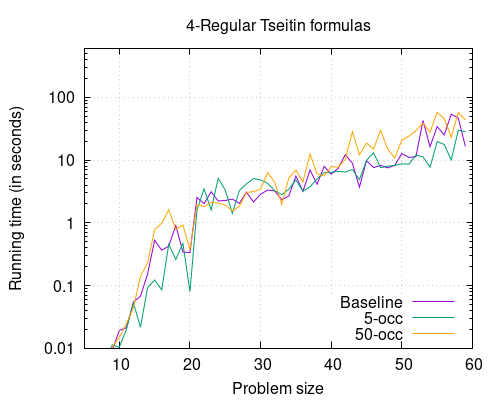}}
      {\includegraphics[width=0.32\linewidth]{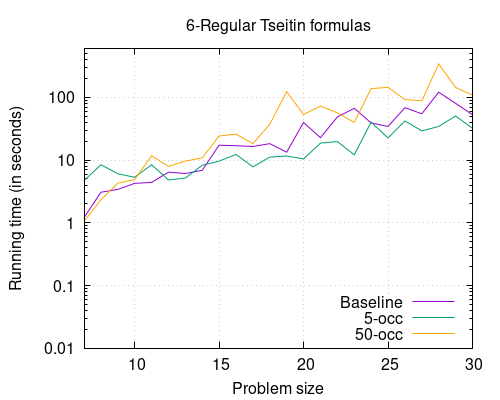}}
      {\includegraphics[width=0.32\linewidth]{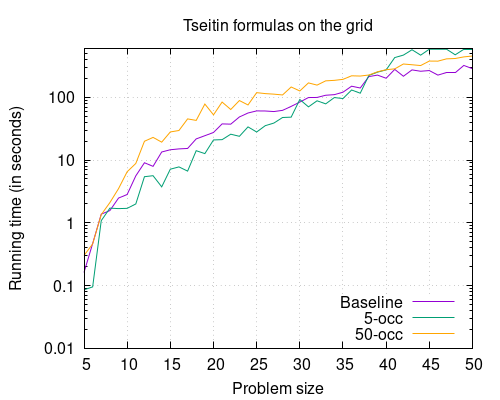}} \\\bigskip
      {\includegraphics[width=0.32\linewidth]{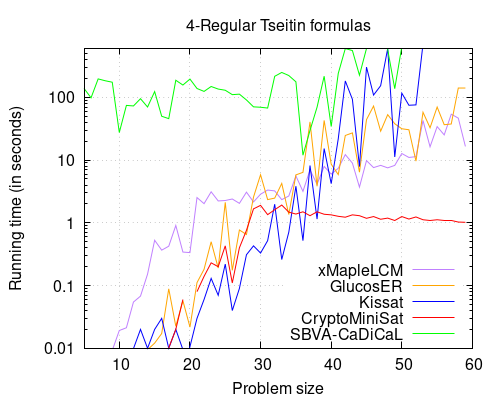}}
      {\includegraphics[width=0.32\linewidth]{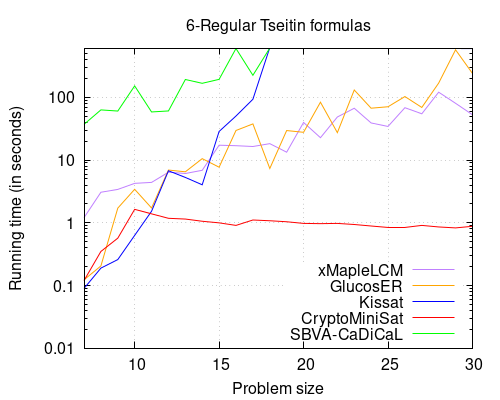}}
      {\includegraphics[width=0.32\linewidth]{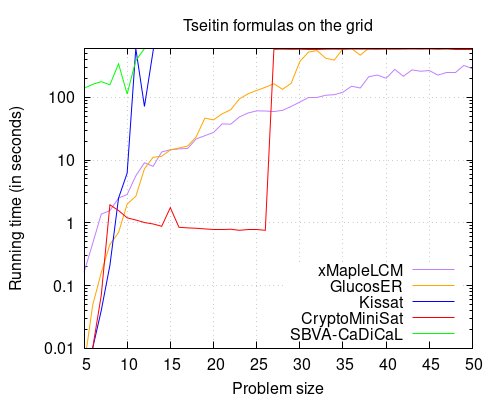}}

      \caption{Performance of different {\xmaple} configurations and state-of-the-art solvers on Tseitin formulas.}       
        %First row compares different choices of DIP. Second row show the impact of only learning the post-DIP clause. Third row shows the effect of changing the threshold that determines how many time a DIP should appear before it is introduced.}
\label{fig:tseitin} 
\end{figure}

The first row of Figure~\ref{fig:tseitin} shows the effect of changing the type of DIP that is used on our {\bf baseline} solver.
A point $(x,y)$ indicates that this particular variant took $y$ seconds (note the logarithmic scale on the $y$
axis) to solve the corresponding Tseitin formula of size $x$. We observe that there is no big difference for regular Tseitin formulas. However, for formulas on the grid, the {\bf middle} DIP is the one that scales best, being the only variant able to solve all instances within the time limit.
In the second row of the same figure, we can see how learning both the pre- and the post-DIP clauses ({\bf baseline} in the plots) is comparable to only learning post-DIP clause ({\bf 1-clause}). Finally, the third row shows that changing the minimum number of times a DIP has to occur before we introduce it has some impact. 50 seems to be too restrictive, whereas 20 ({\bf baseline}) is slightly better than 5, which might end up adding too many DIPs.

The last row of Figure~\ref{fig:tseitin} compares the {\bf baseline} configuration of {\xmaple} with state-of-the-art tools.
The conclusion is clear: our DIP-based {\xmaple} solver is the only one
that can process all instances within the time limit. {\glucoser} scales well on regular formulas, but not on
formulas on the grid. Despite these formulas are in principle trivial for
{\crypto}, since the application of Gaussian elimination in a
preprocessing step solves them, this does not seem to work for large
formulas on the grid. Finally, {\kissat} and {\sbva} perform
poorly, solving only very small instances.

\begin{figure}[t]
\centering
    {\includegraphics[width=0.32\linewidth]{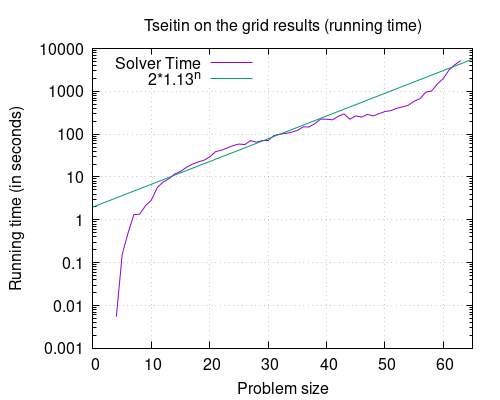}} \qquad
    {\includegraphics[width=0.32\linewidth]{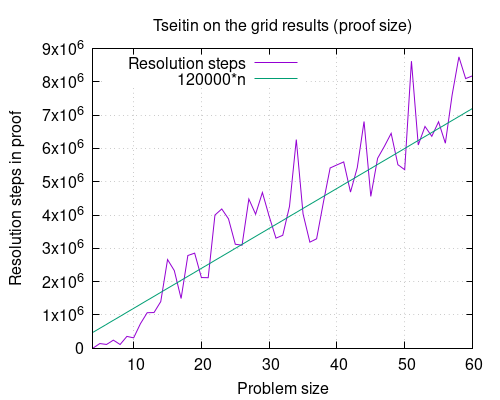}}
\caption{Performance analysis of the baseline DIP-based solver on Tseitin formulas on the grid. The plot on the left represents running time (with logarithmic scale on $y$), whereas the one on the right shows the number of resolution steps in the generated proof.}
\label{fig:tseitin-proof} 
\end{figure}

In order to understand whether the behavior of {\xmaple} was
polynomial for these formulas, we generated more challenging Tseitin
formulas on the grid. Results can be seen in the left plot 
of Figure~\ref{fig:tseitin-proof}, with logarithmic scale
on the $y$ axis.  There is little doubt that the runtime of our solver
ends up being exponential w.r.t. the size of the problem. However, we
went one step further and studied how large the generated trimmed DRAT proofs
were. On the right plot, we show the number of
resolution steps in the proof. Despite having no theoretical support
for that, it seems that although the solver takes exponential time in
finding a proof, its size might be polynomial w.r.t. the problem
size. That would indicate that our problem is using search heuristics
that are not good enough to quickly find a short proof.

\subsection{Xorified Random k-XOR Formulas}

Another family where DIP-based systems perform very well are random
$k$-xor formulas, where xorification~\cite{BenSasson:sizespace} has been applied, i.e.,
replacing variables by xors of fresh variables. We used
CNFgen to obtain 2 sets of formulas: xor constraints of length 3
applying xorification with 2 and 3 variables.
All formulas we consider are unsatisfiable, the
number of variables before xorification is equal to the number of
clauses, and we increase this number to get progressively more
difficult formulas. A time limit of 600 seconds was used.

\begin{figure}[t]
  \centering
      {\includegraphics[width=0.32\linewidth]{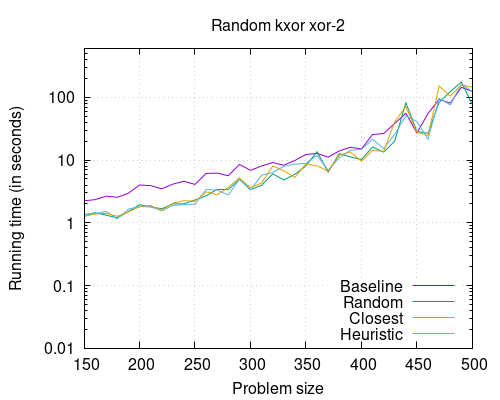}}
      {\includegraphics[width=0.32\linewidth]{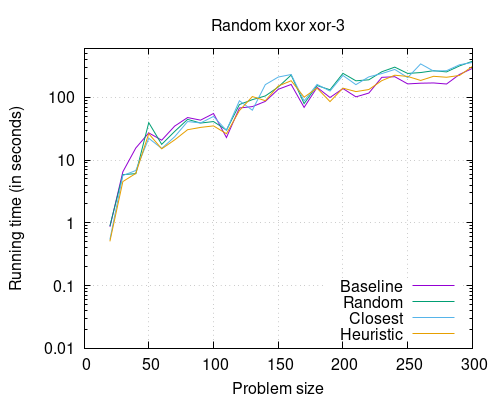}} \\\bigskip\bigskip
      {\includegraphics[width=0.32\linewidth]{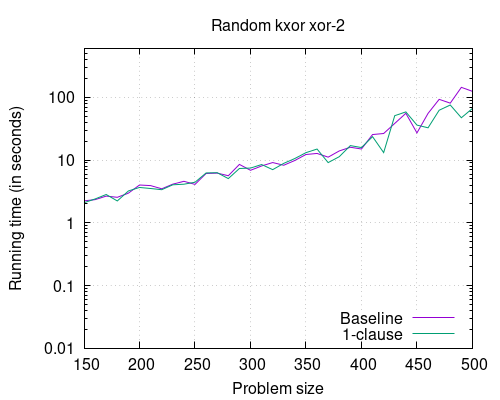}}
      {\includegraphics[width=0.32\linewidth]{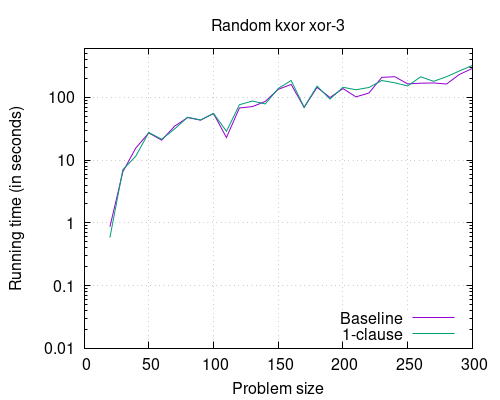}} \\\bigskip\bigskip
      {\includegraphics[width=0.32\linewidth]{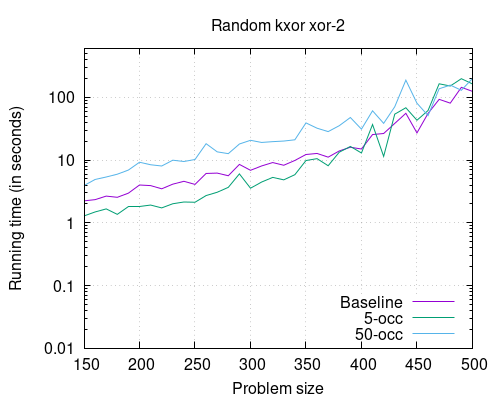}}
      {\includegraphics[width=0.32\linewidth]{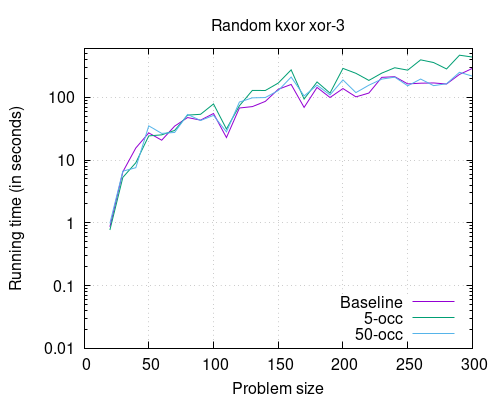}} \\\bigskip\bigskip
      {\includegraphics[width=0.32\linewidth]{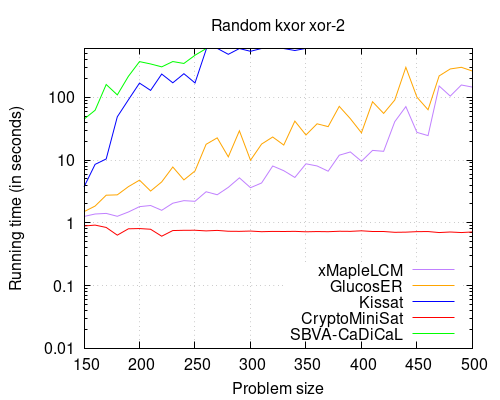}}
      {\includegraphics[width=0.32\linewidth]{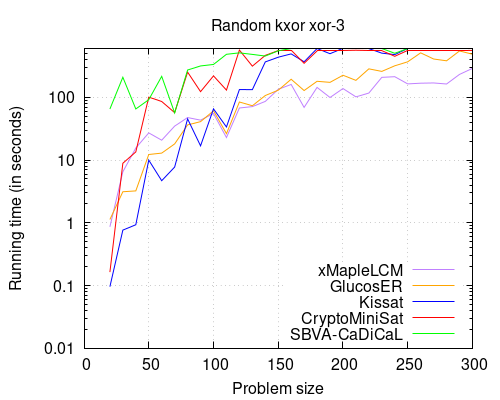}}
      \caption{Performance of different {\xmaple} configurations and state-of-the-arts solvers on XORified rand-kxor formulas.}
\label{fig:xor} 
\end{figure}

The performance of the different {\xmaple} configurations  can be seen
in Figure~\ref{fig:xor}. In the first row, we notice that there is not much difference if
we change the DIP choice. A similar phenomenon can be observed in the second row,
where the use of one or two learned DIP clauses does not have big impact on the performance.
In the third row, we can see that changing the minimum number of occurrences of DIPs does affect the
speed of the system: choosing 20 seems to give more consistent behavior.

Finally, in the last row of Figure~\ref{fig:xor} we notice that {\crypto} can
only benefit from its preprocessing step when xorification with 2
variables is applied. Our {\xmaple} solver is the one the performs
best with {\glucoser} in second place. Finally, both {\sbva} and
{\kissat} are only able to solve small instances.

%% \begin{figure}[t]
%%   \centering
%%       {\includegraphics[width=0.45\linewidth]{experiments/results-journal-DIP/final/randkxor/rand-or-2-dip.png}}
%%       {\includegraphics[width=0.45\linewidth]{experiments/results-journal-DIP/final/randkxor/rand-or-3-dip.png}} \\\bigskip\bigskip
%%       {\includegraphics[width=0.45\linewidth]{experiments/results-journal-DIP/final/randkxor/rand-or-2-1clause.png}}
%%       {\includegraphics[width=0.45\linewidth]{experiments/results-journal-DIP/final/randkxor/rand-or-3-1clause.png}} \\\bigskip\bigskip
%%       {\includegraphics[width=0.45\linewidth]{experiments/results-journal-DIP/final/randkxor/rand-or-2-occ.png}}
%%       {\includegraphics[width=0.45\linewidth]{experiments/results-journal-DIP/final/randkxor/rand-or-3-occ.png}}
%%       \caption{Performance of different {\xmaple} configurations on ORified rand-kxor formulas}
%% \label{fig:or-dip}
%% \end{figure}

%% \begin{figure}[t]
%%   \centering
%%       %% {\includegraphics[width=0.45\linewidth]{experiments/results-journal-DIP/final/randkxor/rand-or-2-others.png}}
%%       %% {\includegraphics[width=0.45\linewidth]{experiments/results-journal-DIP/final/randkxor/rand-or-3-others.png}} \\\bigskip\bigskip
%%       {\includegraphics[width=0.32\linewidth]{experiments/results-journal-DIP/final/randkxor/rand-xor-2-others.png}}
%%       {\includegraphics[width=0.32\linewidth]{experiments/results-journal-DIP/final/randkxor/rand-xor-3-others.png}}

%%       \caption{Performance of state-of-the-art tools on XORified rand-kxor formulas.}
%% \label{fig:xor-others} 
%% \end{figure}

\subsection{Matching of Properly Intersecting Intervals}

We are given a sequence of numbers $(a_1,\ldots,a_n)$, initially all
set to zero, and a set of operations, each consisting of assigning a
certain number to a contiguous subsequence of the $a_i$'s, defined
by an interval. The goal is to perform each operation exactly once while
maximizing the number of pairs of consecutive numbers $(a_i,a_{i+1})$
that are different at the end of the process. Given some additional
conditions, this can be formulated as maximum bipartite
matching problem on a certain graph. By removing some edges from this
graph, an unsatisfiable problem is generated. A more detailed
description can be found in~\cite{SATcomp2023}.

We downloaded the 23 instances submitted to the SAT Competition, using
the Global Benchmark Database~\cite{GlobalBenchmarkDatabase}, and
observed that no system without extended-resolution reasoning could
solve any benchmark in a time limit of 2 hours. We used a much larger
timeout than with the two previous families because for this family we
are not able to construct increasingly larger benchmarks and see how
the solvers scale.

Table~\ref{tab:intervals} reports on our experimental evaluation. In
the first block we can see that the {\bf random} and {\bf heuristic} DIP
choices do significantly improve upon our {\bf baseline} solver. The
second block row shows how the behavior of the {\bf heuristic} configuration
can be further improved by increasing the threshold of minimum DIP
occurrences to 50.  On the contrary, a smaller threshold of 5
significantly degrades the performance, whereas only learning the post-DIP clause
slightly slows it down. Finally, the third block compares our best {\xmaple} configuration
with state-of-the-art solvers. 
{\xmaple} and {\glucoser} are the only systems
able to process any instance within the given time limit. Hence, the use of extended resolution
is crucial in this set of benchmarks. Our comparison shows that {\glucoser} and {\xmaple} solve
the same number of instances, the former being only slightly faster.

  \begin{table}[t]
  \begin{center}
    \small
    \begin{tabular}{|c|c|c|c|c|c|}\cline{2-5}
      \multicolumn{1}{c|}{\bf }     & {\bf Baseline} & {\bf Random} & {\bf Closest} & {\bf Heuristic} \\\cline{1-5}
      SOLVED                        & 6            & {\bf 13}     & 10            & {\bf 13}  \\\cline{1-5}
%      BEST                          & 0            & 4            & 2             & {\bf 8}  \\\cline{1-5}
      Average (secs.)               & 6011         & 4337         & 4818          & {\bf 4220} \\\cline{1-5}
      \multicolumn{5}{c}{}\\\cline{2-5}
      \multicolumn{1}{c|}{\bf }     & {\bf Heuristic} & {\bf 1-clause} & {\bf Occ-5} & {\bf Occ-50} \\\cline{1-5}
      SOLVED                        & 13              & 11           & 6             & {\bf 14}        \\\cline{1-5}
%      BEST                          & 3               & 3            & 1             & {\bf 7}        \\\cline{1-5}
      Average (secs.)               & 4220            & 4995         & 6057          & {\bf 3913}      \\\cline{1-5}
      \multicolumn{5}{c}{}\\\hline
      \multicolumn{1}{c|}{\bf }     & {\bf Occ-50}    & {\bf {\glucoser}} & {\bf {\kissat}} & {\bf {\crypto}} & {\bf {\sbva}} \\\hline
      SOLVED                        & {\bf 14}        & {\bf 14}           & 0             & 0     & 0    \\\hline
%      BEST                          & 4               & {\bf 11}           & 0             & 0     & 0      \\\hline
      Average (secs.)               & 3913            & {\bf 3622}         & 7200          & 7200  & 7200     \\\hline

    \end{tabular}

  \end{center}
  \caption{Average running times in seconds of DIP-learning variants on 
  properly intersecting intervals formulas. The best performing systems are in bold.}
  \label{tab:intervals}
\end{table}

  \subsection{Performance Analysis on SAT Competition Benchmarks}

  Our experimental evaluation concludes with the lessons we have
  learned from executing our DIP implementation on the SAT competition
  benchmarks.  We used benchmarks from the 2023 and 2024 editions. In
  the latter, the number of instances from the Tseitin, Xorified
  random $k$-xor and Interval families we have described is much lower
  than in the 2023 edition, and hence we present results from the 2023
  competition in order to show the possible gains or losses of adding
  DIP-based reasoning to solvers. We first report on the overhead
  caused by the DIP detection algorithm and the subsequent additional
  work to retrieve the clauses to be learned. For our baseline
  DIP-based system, where two clauses are learned and hence is the
  most computationally demanding method, in 4.5\% of the benchmarks
  the DIP-related work represented between 10 and 15\% of the total
  runtime; in 11.5\% of the benchmarks between 5 and 10\%; in 35.5\%
  between 2.5 and in 48.5\% less than 2.5\%. These data show that DIP
  computation does not significantly slow down the solver. Another
  interesting information concerns the percentage of conflicts where
  there is at least one DIP, which was on average 63\%. This implies
  that we do need to have filtering mechanisms to discard some of
  them. Otherwise, the search would be totally dominated by DIPs.

\begin{figure}[t]
\centering
    {\includegraphics[width=0.45\linewidth]{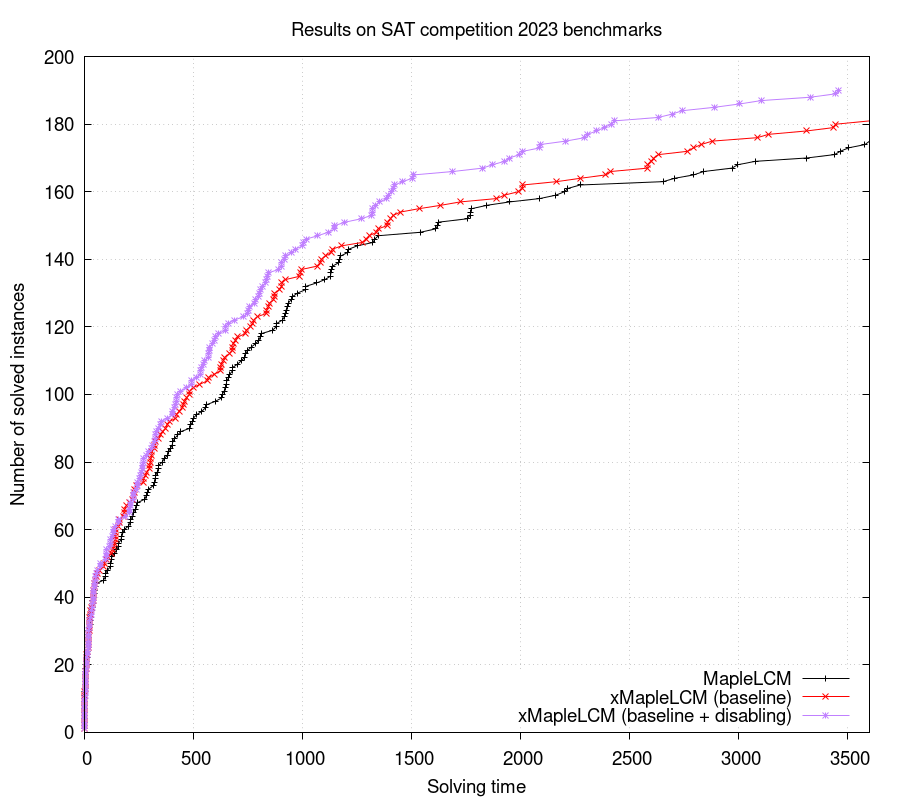}}
    {\includegraphics[width=0.45\linewidth]{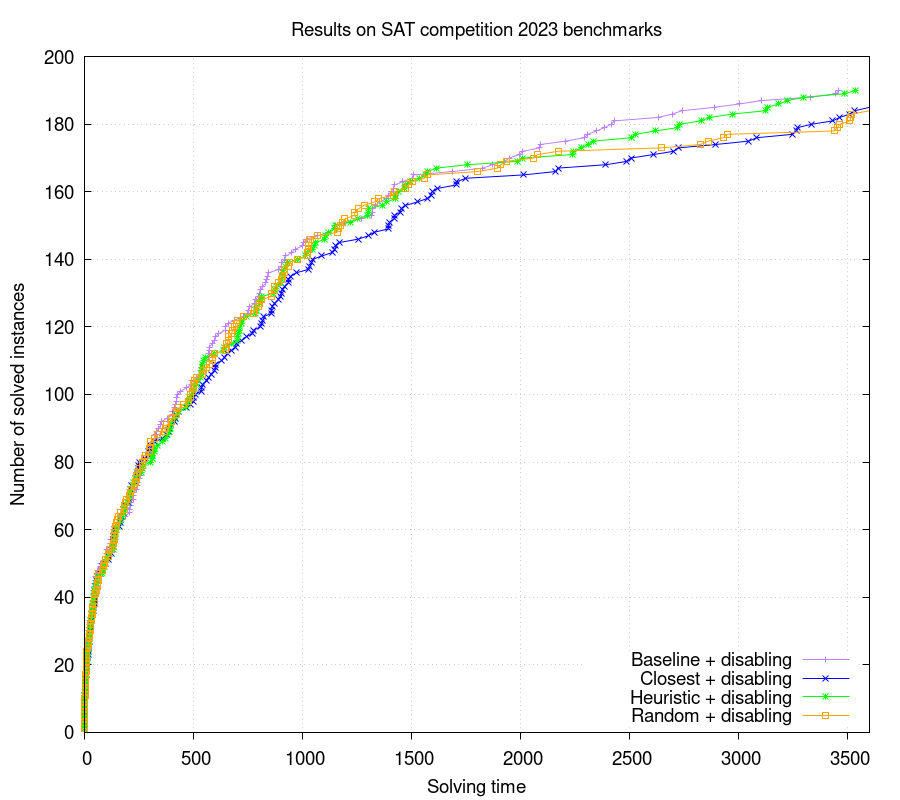}}
\caption{Performance of DIP variants on 2023 SAT Competition benchmarks.}
\label{fig:sat-comp} 
\end{figure}

  In the left CDF plot of Figure~\ref{fig:sat-comp} we can see that
  our baseline DIP-based {\xmaple} solver outperforms the CDCL-based {\maple}
  that it is based on.  When we analyzed the concrete benchmarks where
  our DIP-based solver outperforms {\maple} and other CDCL-based
  solvers like {\kissat}, we observed that for all of them
  the percentage of decisions on extended variables by {\xmaple} is  quite
  large (over 10\% of the decisions), whereas for the remaining 
  benchmarks it is very low (in 70\% of the benchmarks less than 1\% of
  decisions are on extended variables). Moreover, this happens no matter
  how many initial variables there are, even though in problems
  with few initial variables
  the introduction of a few extended variables could potentially quickly
  dominate the decision heuristic.

  This is  important since it shows that already existing decision
  heuristics like VSIDS or LRB somehow infer
  whether the newly introduced variables improve the behavior of
  the system.  This is why we implemented, on top of our baseline DIP
  system, a procedure that computes the percentage of decisions on
  extended variables. If after a fixed number of conflicts it is
  still lower than 3\%, it discontinues
  DIP learning and performs 1-UIP learning from that moment
  onwards. This is a very preliminary step in the direction of trying to
  automate the decision of whether to apply DIP reasoning or not, but
  the outcome, which is found on the left CDF plot of
  Figure~\ref{fig:sat-comp}, is very positive, showing that 
  a solver disabling DIP reasoning ({\xmaple}-disabling in the plot) 
  when the percentage of decisions on extended variables is low
  is still able to solve Tseitin, random $k$-xor and Interval benchmarks, 
  but the overall performance
  is not significantly worsened on the remaining instances. Finally, 
  the right CDF plot shows the impact of changing the type of DIP 
  into the disabling solver: the {\bf heuristic} and {\bf middle} DIPs, the latter being used in the baseline configuration, 
  are much better than {\bf random} and {\bf closest} ones.
   
\bigskip
\noindent
    {\bf Learned lessons from experimental evaluation.} Let us finish
    this section by pointing out the main lessons we have learned from this
    experimental evaluation.

    \begin{itemize}
    \item
      We have identified three families of benchmarks for which the
      dynamic addition of extension variables improves upon CDCL in a
      systematic manner. This is an improvement w.r.t. the results in
      the {\glucoser} paper~\cite{Audemard2010ARO}, where a few
      initial benchmarks were identified but it was not proved that
      extended-resolution methods always outperform CDCL on increasingly
      more complex instances.
      \smallskip
    \item
      We have proved that, even though the way extended variables are
      identified is completely different in {\xmaple} and {\glucoser},
      they both behave well on the same instances. 
      On the other hand,
      our preliminary experiments revealed that randomly
      choosing arbitrary pairs of (non-DIP) literals to create
      definitions does not provide any speedup w.r.t. CDCL. Hence, 
      both {\glucoser} and {\xmaple} are able to identify meaningful semantic properties about
      concrete literals that make them suitable for creating an
      extended variable definition.  The data we collected
        show that {\glucoser} typically introduces around four times more
        extension variables than {\xmaple} and that the extended
        variables generated by {\glucoser} and {\xmaple} are neither disjoint nor totally overlap. Only about
        25\% of the definitions over original variables
        introduced in {\xmaple} are also introduced in {\glucoser}.
      \smallskip
    \item
      We have learned that in most of the families where extended
      resolution works there is a large xor component. Additionally,
      when this is xor information is somehow hidden in the formula,
      our dynamic approach outperforms static methods like the ones in
      {\crypto} and {\sbva}.
      \smallskip
    \item
      Our ablation experiments have shown that the most important design choice
      is the one of determining which DIP is to be chosen. Among all the possibilities
      we have devised, {\bf middle} and {\bf heuristic} are the preferable ones.
      \smallskip
    \item
      Finally, we have noticed that by inspecting the percentage of decisions
      on extended variables, one can detect whether our DIP-based is helping the solver.
      This has allowed us to further improve the behavior of our solver on sets of benchmarks
      where only a few of them are advantageous for extended resolution.
    \end{itemize}

\section{Conclusions and Future Work}
\label{sec:conclusions}

We introduce a novel extended resolution clause learning (ERCL)
algorithm that, when implemented on top of a CDCL solver, turns out to
be beneficial for a variety of problems, in particular Tseitin, random
$k$-xor and interval matching formulas.
We view this as a step towards more effective
methods for incorporating extended resolution to form more powerful proof systems.
We show that the only previously existing attempt to
incorporate ER into CDCL performs reasonably well on the same
instances.
Hence, we have identified classes of formulas for which
extended resolution can be automated in at least two different ways
and still outperform CDCL. Considering the different nature of the two
methods, this  deserves further study. Further, we also
introduce a new heuristic that allows our ERCL solver xMapleSAT to
perform similarly to the baseline CDCL solver MapleLCM, thus being
able to get the best of both worlds, i.e., the benefit of ERCL,
without sacrificing performance of the CDCL solver on, say, SAT
Competition Main Track 2023 instances.

As future work, we plan to investigate a variety of machine learning
based heuristics (e.g., branching) specialized for the ERCL
method. Also, the use of $k$-IPs with $k>2$ is part of our next
steps. Finally, we intend to evaluate the impact of our
  technique on benchmarks from approximate hash-based counting, for
  which XOR reasoning is crucial. On the theoretical side,
challenging questions like determining whether DIP- or $k$-IP based ER
simulates unrestricted ER are going to be central to our research
efforts.

\bigskip

\noindent
{\bf Acknowledgments.} 
We thank the anonymous referees for helpful comments and corrections.

%Future work:
%
%    * K-IP generalization?
%    
%    * Heuristics for branching?

%    * Everything is "immature". A lew line of research has been opened.

%    * Understand how "polynomial" proofs for Tseitin are found

%More questions/future work:

%* Does DIP learning simulate full ER?

%* Give an explicit construction of how DIP learning gives poly size refutations for Tseitin tautologies.

%* Extend DIP / {\glucoser} to be advantageous on a wider range of principles, even the general SAT competition.

%% %%
%% %% Bibliography
%% %%

%% %% Please use bibtex, 

\bibliographystyle{alphaurl}% the mandatory bibstyle
\bibliography{paper,logic}

%\bibliography{paper,logic}

%% \appendix

\end{document}